\documentclass[smallextended,numbook]{myownsvjour3}
\usepackage{graphicx}

\newcommand{\beq}{\begin{equation}}
\newcommand{\beqa}{\begin{eqnarray}}
\newcommand{\eeq}{\end{equation}}
\newcommand{\eeqa}{\end{eqnarray}}
\newcommand{\abs}[1]{\left\vert#1\right\vert}

\newcommand{\braket}[2]{\left<#1\vert#2\right>}
\newcommand{\cas}{\noindent $\bullet$ {\hskip 3pt}}
\renewcommand{\d}{{\rm d}}
\newcommand{\dif}[2]{{\frac{\d #1}{\d #2}}}
\newcommand{\e}{{\rm e}}
\newcommand{\eps}{\varepsilon}
\newcommand{\eo}{{(\pm)}}
\newcommand{\even}{{(+)}}
\newcommand{\frad}[2]{\displaystyle{\displaystyle#1\over\displaystyle#2}}
\newcommand{\g}{\gamma}
\newcommand{\h}{\widehat}
\newcommand{\ii}{{\rm i}}
\newcommand{\ket}[1]{\left.\vert#1\right>}
\renewcommand{\max}{{\rm max}}
\renewcommand{\min}{{\rm min}}

\newcommand{\odd}{{(-)}}
\newcommand{\p}{\psi}
\newcommand{\pbar}{\overline{\psi}}

\newcommand{\m}{{\bf m}}
\newcommand{\n}{{\bf n}}
\renewcommand{\Im}{\mathop{\rm Im}\nolimits}
\renewcommand{\P}{\Pi}
\renewcommand{\Re}{\mathop{\rm Re}\nolimits}

\journalname{Journal of Statistical Physics}

\begin{document}

\title{Survival of classical and quantum particles in the presence of traps}

\author{P.L. Krapivsky \and J.M. Luck \and K. Mallick}

\institute{
P.L. Krapivsky
\at
Department of Physics, Boston University, Boston, MA 02215, USA\\
Institut de Physique Th\'eorique, CEA Saclay and CNRS URA 2306,
91191 Gif-sur-Yvette cedex, France\\
\email{pkrapivsky@gmail.com}
\and
J.M. Luck and K. Mallick
\at
Institut de Physique Th\'eorique, CEA Saclay and CNRS URA 2306,
91191 Gif-sur-Yvette cedex, France\\
\email{jean-marc.luck@cea.fr,kirone.mallick@cea.fr}
}

\date{\today}

\maketitle

\begin{abstract}
We present a detailed comparison of the motion
of a classical and of a quantum particle in the presence of trapping sites,
within the framework of continuous-time classical and quantum random walk.
The main emphasis is on the qualitative differences
in the temporal behavior of the survival probabilities of both kinds of particles.
As a general rule,
static traps are far less efficient
to absorb quantum particles than classical ones.
Several lattice geometries are successively considered:
an infinite chain with a single trap,
a finite ring with a single trap,
a finite ring with several traps,
and an infinite chain and a higher-dimensional lattice
with a random distribution of traps with a given density.
For the latter disordered systems,
the classical and the quantum survival probabilities
obey a stretched exponential asymptotic decay, albeit with different exponents.
These results confirm earlier predictions,
and the corresponding amplitudes are evaluated.
In the one-dimensional geometry of the infinite chain,
we obtain a full analytical prediction for the amplitude of the quantum problem,
including its dependence on the trap density and strength.

\end{abstract}

\newpage

\section{Introduction}
\label{intro}

A highly important quantity in the theory of random walk
is the number of distinct sites visited by the walk in a given time~\cite{mw}.
The distribution of this number of sites
governs the behavior of many physical processes.
The trapping of diffusing particles is
the most famous example~\cite{BV,Varadhan1,Varadhan2}.
It has been shown by Donsker and Varadhan~\cite{Varadhan1,Varadhan2}
that the survival probability $P(t)$ of a classical particle diffusing in an environment
containing randomly placed static traps can be expressed
in terms of the distribution of distinct sites visited by the walk
in the absence of traps.
Classical trapping models of this kind have been explored
thoroughly~\cite{Grassberger,Wilczek,Klafter,Lebowitz1,Lebowitz2,BRW}.
They have received many applications, including chemical transport,
electron-hole recombination, exciton trapping and annihilation
(see~\cite{hk,DanyRW,PaulKBook} for reviews).
The long-time asymptotics of the survival probability of a classical walker
in the presence of a random distribution of traps has been established
by Balagurov and Vaks~\cite{BV},
and more rigorously by Donsker and Varadhan~\cite{Varadhan1,Varadhan2}.
In dimension~$d$ it displays a stretched exponential behavior of the form
\beq
P(t)\sim\exp\left(-A_d\,t^{d/(d+2)}\right),
\label{pc}
\eeq
where the amplitude $A_d$ is known exactly~(see~(\ref{a1res}),~(\ref{adres})).
The above result can indeed be related to the form of the Lifshitz tail
of the density of states of the appropriate random operator~\cite{lif,lgp,fp}.
The slower-than-exponential decay of the survival probability
is caused by the existence of arbitrarily large trap-free regions,
where the particle can survive for an anomalously large time.
At short times, the survival probability however decreases exponentially,
according to the Rosenstock formula,
which is the result of a mean-field analysis~\cite{Rosenstock}.
Although the survival probability has been studied in great detail
for a whole breadth of trapping models,
some problems still remain open.
In particular, the crossover between the small-time (Rosenstock)
and large-time (Lifshitz) regimes is not fully understood.
The scaling properties in the crossover region
have been investigated in detail by means of Monte-Carlo simulations~\cite{Barkema}.
On the theoretical side,
a precise calculation of the survival probability throughout the Lifshitz regime
is only possible in one dimension.
In higher dimensions,
going beyond the leading term given in~(\ref{pc})
requires sophisticated field-theoretical techniques
involving instantons~\cite{inst1,inst2,inst3,inst4,NLEPL,inst5,inst6}.

At low temperatures, quantum effects eventually become relevant,
overshadowing the diffusive nature of transport.
In the presence of traps,
the asymptotic form~(\ref{pc}) of the survival probability
is accordingly expected to be modified in a quantum-mechanical setup.
The long-time behavior of the survival of a quantum particle
in the presence of traps was first studied in a series of works
by Parris~\cite{P1,P2,P3,ParrisJSP}.
The motivation for these investigations was to understand
the low-temperature reaction dynamics and exciton quenching by isolated impurities
in the coherent regime.
The effective dynamics for the reduced density matrix of the exciton
was shown to be given by a non-Hermitian tight-binding model,
with an imaginary optical potential
representing scattering, dephasing and absorption
by the impurities~\cite{RMP,Lindenberg,Huber,Silbey}.
The long-time asymptotics of the survival probability $\P(t)$
of a quantum exciton hopping in a medium with a finite density of traps
was found to be
\beq
\P(t)\sim\exp\left(-B_d\,t^{d/(d+3)}\right).
\label{pq}
\eeq
In contrast with the result~(\ref{pc}) pertaining to the classical case,
the dependence of the amplitude $B_d$
on the density of traps and other parameters
could only be worked out within approximations~\cite{P1,P2,P3,ParrisJSP}.
One of the main motivations for undertaking the present work
was to obtain a better understanding of the amplitude $B_d$
(see the results~(\ref{b1res}),~(\ref{bdres})).

In more modern language, the motion of the exciton can be interpreted
as a {\it quantum walk}.
Quantum analogs of classical random walks have been defined in
two pioneering papers~\cite{ADZ,FG} as a useful concept to implement original algorithms
in quantum information theory.
Due to interference effects,
the properties of quantum walks can drastically differ from their classical counterparts.
Clever implementations of quantum algorithms
can thus lead to much faster computations~\cite{Goldstone,ChildsPRL,AC}
(see~\cite{Kempe,Ambainis} for reviews).
As far as experimental realizations of quantum walks are concerned,
it is worth mentioning the promising area of photon propagation
in waveguide lattices~\cite{perets,peruzzo}.
On the theoretical side,
two distinct classes of quantum walks have been studied.
In discrete-time quantum walks~\cite{ADZ,NK,MBT,TFM},
in addition to its position,
the particle has a discrete internal quantum degree of freedom (a `quantum coin'),
which experiences unitary evolution at each step.
In continuous-time quantum walks~\cite{FG,TL,benAvraham},
no internal state is involved,
and the role of the quantum Hamiltonian
is played by some hopping matrix,
such as the adjacency matrix on the underlying lattice or graph~\cite{Eric}.
These two classes of models can be reconciled
by means of a precise limiting procedure~\cite{FS},
which turns out to be more subtle than its classical counterpart,
namely the emergence of diffusion as the continuum limit
of discrete random walks (see~\cite{qwrev} for a review).

Last but not least, there has been considerable progress lately
in manipulating quantum-mechanical systems.
The effect of decoherence in low-dimensional quantum systems
can thus be investigated experimentally~\cite{Haroche}.
A pa\-ra\-di\-g\-m for the loss of coherence
in quantum transport is provided by quantum walks in the presence of traps.
The recent years have witnessed an upsurge of interest in this kind
of models~\cite{KNS,MBA,MPB,AMB,Ag,MulkenRep,GAS,Gonulol,Anish}.

The goal of the present work is to present a systematic comparison
of the motion of a classical and of a quantum particle in the presence of traps,
within the framework of the continuous-time classical and quantum random walk.
The main emphasis will be put on the temporal behavior
of the corresponding survival probabilities $P(t)$ and $\Pi(t)$.
A common feature of all the situations to be considered in this work
is that quantum particles are able to benefit from interference effects
in order to efficiently avoid the traps.
As a consequence, static traps are far less efficient
to absorb quantum particles than classical ones.
This qualitative difference will manifest itself in various settings.

The outline of this article is as follows.
In section~\ref{warm} we review the basic features of the classical and
quantum motion of a free particle on the one-dimensional chain,
emphasizing the most salient differences between both types of dynamics.
Section~\ref{chain} is devoted to the situation of an infinite chain with a single trap.
Whereas classical random walk is recurrent,
with a survival property decaying as $P(t)\sim t^{-1/2}$,
the quantum particle has a non-vanishing probability to escape ballistically
to infinity, and thus to survive forever.
The asymptotic survival probability $\P_\infty$
is calculated for different initial conditions.
It is shown to display generically
a {\it non-monotonic} behavior with respect to the trapping strength.
The geometry of a finite ring of $N$ sites
with a single trap is studied in Section~\ref{ring}.
The classical survival probability decays exponentially,
with the associated characteristic time scaling diffusively, i.e., as $N^2$.
The dynamics of the quantum particle demonstrates two main differences
with respect to~the classical case.
The survival probability tends to a non-zero asymptotic value,
which is generically $\P_\infty=1/2$ in this geometry.
The relaxation time characterizing the convergence toward that value
grows asymptotically as $N^3$~\cite{P1}.
This characteristic time is calculated in various regimes
and is shown to exhibit again a non-monotonic dependence on the trapping strength.
If several traps are present on a ring,
the determination of the {\it avoiding modes}
which are responsible for the non-vanishing
of the asymptotic survival probability $\P_\infty$,
leads to an elegant problem of combinatorics,
which we solve in Section~\ref{rings}.
The case of the chain with a fixed density of traps
is studied in Section~\ref{1d}.
In the classical situation,
we review the predictions of the theory of Lifshitz tails.
We thus introduce tools from the theory
of one-dimensional disordered systems~\cite{NL,JMLivre},
which are also pertinent to the analysis of the quantum situation.
In the latter case, for a large but finite system,
the decay rate of the survival probability is a fluctuating quantity,
whose distribution is investigated.
We are thus able to determine exactly
the amplitude $B_1$ of the stretched exponential~(\ref{pq}),
including its dependence on the density $c$ of traps
and on the trapping strength $\g$ (see~(\ref{b1res})).
The problem of a classical and of a quantum particle diffusing
on a higher-dimensional lattice with traps is considered in Section~\ref{highd}.
Here, an exact calculation of the decay rate seems to be out of reach.
We are led to make an Ansatz for its scaling behavior,
and draw conclusions on the behavior of the survival probability~$\Pi(t)$
(see~(\ref{bdres})).
The last section is devoted to concluding remarks.

\section{A warming up: classical and quantum diffusion on the chain}
\label{warm}

In this section we recall the basic differences between
a classical and a quantum particle diffusing on the infinite chain.

\subsection{A classical particle}
\label{warmc}

Consider a classical particle performing continuous-time random walk on the chain.
Here and throughout the following, we use dimensionless variables,
assuming that the lattice spacing and the hopping frequency are set equal to unity.
The probability $p_n(t)$ for the particle to be at site $n$ at time $t$
obeys the differential equation
\beq
\dif{p_n(t)}{t}=p_{n+1}(t)+p_{n-1}(t)-2p_n(t).
\label{wcdif}
\eeq
Let us assume that the particle is launched at the origin at time $t=0$.
Introducing the momentum $q$ conjugate to $n$,
we readily obtain
\beq
\h p(q,t)=\sum_n p_n(t)\,\e^{-\ii nq}=\e^{-2(1-\cos q)t}.
\eeq
The probabilities $p_n(t)$ therefore read
\beq
p_n(t)=\e^{-2t}I_n(2t),
\label{cprob}
\eeq
where the $I_n$ are the modified Bessel functions.
The bulk of the probability profile has a Gaussian form
\beq
p_n(t)\approx\frac{\e^{-n^2/(4t)}}{\sqrt{4\pi t}},
\eeq
characteristic of a diffusive motion.
In particular, the return probability
\beq
p_0(t)=\e^{-2t}I_0(2t)\approx\frac{1}{\sqrt{4\pi t}}
\label{cret}
\eeq
decays monotonically to zero.

\subsection{A quantum particle}
\label{warmq}

Consider now a quantum particle propagating coherently along the chain.
We again use dimensionless variables, assuming that Planck's constant,
the lattice spacing and the hopping amplitude are set equal to unity.
Within this framework,
the wavefunction $\p_n(t)$ of the particle at site~$n$ at time~$t$
obeys the time-dependent reduced tight-binding equation
\beq
\ii\,\dif{\p_n(t)}{t}=\p_{n+1}(t)+\p_{n-1}(t).
\label{wqdif}
\eeq
Let us again assume that the particle is launched at the origin at time $t=0$.

The dispersion law between energy $\eps$ and momentum $q$ reads
\beq
\eps=2\cos q.
\label{qdisp}
\eeq
The group velocity is therefore $v=\d\eps/\d q=-2\sin q$.
We have
\beq
\h\p(q,t)=\e^{-2\ii\cos q t},\qquad\p_n(t)=\ii^{-n}\,J_n(2t),
\eeq
and so the quantum probabilities read
\beq
\abs{\p_n(t)}^2=(J_n(2t))^2,
\label{qprob}
\eeq
where the $J_n$ are the Bessel functions.
These probabilities take appreciable values in the range $-2t<n<2t$,
which spreads ballistically with the maximal velocity $v_\max=2$.
The probability profile exhibits sharp ballistic fronts near $n=\pm2t$,
whose width scales as $t^{1/3}$~\cite{TL},
and it decays exponentially beyond these fronts.
The return probability decays~as
\beq
\abs{\p_0(t)}^2=(J_0(2t))^2\approx\frac{\cos^2(2t-\pi/4)}{\pi t}.
\label{qret}
\eeq

\begin{figure}[!ht]
\begin{center}
\includegraphics[angle=-90,width=.55\linewidth]{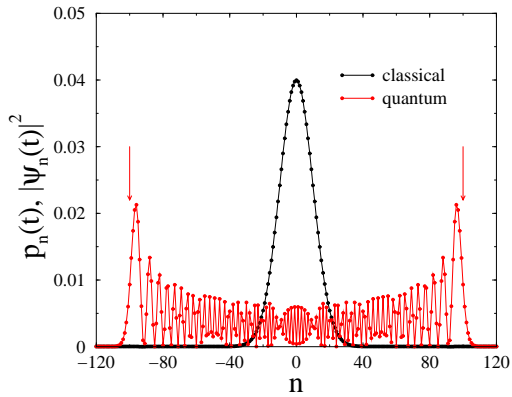}
\caption{\small
Probability profiles $p_n(t)$
(see~(\ref{cprob})) and $\abs{\p_n(t)}^2$ (see~(\ref{qprob}))
of a classical and a quantum particle launched at the origin
against the particle's position $n$, for $t=50$.
Arrows: nominal positions of the ballistic fronts
in the quantum case ($n=\pm100$)
(Color online).}
\label{profil}
\end{center}
\end{figure}

\begin{figure}[!ht]
\begin{center}
\includegraphics[angle=-90,width=.55\linewidth]{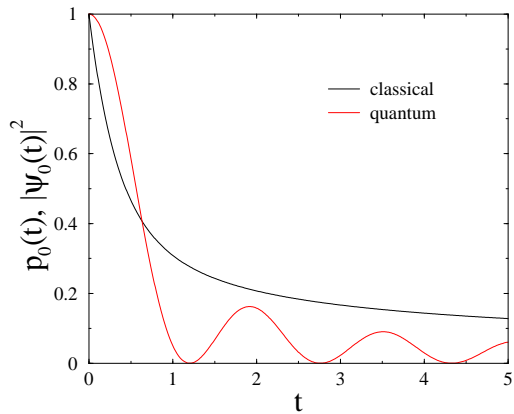}
\caption{\small
Return probabilities $p_0(t)$ (monotonic, see~(\ref{cret}))
and $\abs{\p_0(t)}^2$ (oscillating, see~(\ref{qret}))
of a classical and a quantum particle, against time $t$
(Color online).}
\label{retour}
\end{center}
\end{figure}

Figures~\ref{profil} and~\ref{retour} illustrate the most prominent differences
between classical and quantum diffusion.
Figure~\ref{profil} shows a plot of both probability profiles for $t=50$.
The classical profile $p_n(t)$ is rather narrow and approximately Gaussian,
whereas the quantum profile $\abs{\p_n(t)}^2$ is much broader and irregular.
The arrows show the nominal positions of the ballistic fronts ($n=\pm100$).
Figure~\ref{retour} shows the return probabilities against time $t$.
The classical probability $p_0(t)$ falls off monotonically and rather slowly,
whereas the quantum one, $\abs{\p_0(t)}^2$, falls off faster on average
and shows infinitely many oscillations.

\section{The infinite chain with a single trap}
\label{chain}

We now study the effect of a single trap
on the motion of a classical and a quantum particle on the infinite chain.

\subsection{A classical particle}
\label{chainc}

Consider a classical particle diffusing
on the infinite chain, with a single trap at the origin.
The strength $\g$ of a classical trap is defined as its absorption rate per unit time.
The differential equation~(\ref{wcdif}) becomes
\beq
\dif{p_n(t)}{t}=p_{n+1}(t)+p_{n-1}(t)-2p_n(t)-\g\delta_{n0}\,p_0(t).
\label{cdif}
\eeq
Let us assume that the particle is launched at site $a\ge0$ at time $t=0$.

We are mostly interested in the survival probability
of the particle up to time $t$:
\beq
P(t)=\sum_np_n(t)=1-\g\!\int_0^tp_0(t')\,\d t'.
\label{csres}
\eeq
Introducing a Laplace variable $s$ conjugate to $t$
and a Fourier variable (momentum) $q$ conjugate to $n$, we obtain
\beq
\h p(q,s)=\frac{\e^{-\ii qa}-\g\,\h p_0(s)}{s+2(1-\cos q)}.
\label{pflc}
\eeq
The quantity $\h p_0(s)$ obeys the self-consistency condition
\beq
\h p_0(s)=\int_0^{2\pi}\frac{\d q}{2\pi}\,\h p(q,s)
=\frac{z_1^a-\g\,\h p_0(s)}{z_2-z_1},
\eeq
where $z_1$ and $z_2$ are the zeros of the denominator of~(\ref{pflc})
in the variable $z=\e^{-\ii q}$ such that $\abs{z_1}<1<\abs{z_2}$, i.e.,
\beq
z_{1,2}=\frac{s+2\mp\sqrt{s(s+4)}}{2}.
\eeq
Hence
\beq
\h p_0(s)=\frac{z_1^a}{\g+z_2-z_1}.
\label{p0lc}
\eeq
The survival probability therefore reads in Laplace space
\beq
\h P(s)=\h p(0,s)
=\frac{1}{s}\left(1-\frac{\g\,z_1^a}{\g+z_2-z_1}\right).
\label{scres}
\eeq
The short-time behavior of the survival probability
is governed by the behavior of~(\ref{scres}) as $s\to\infty$.
We have
\beq
\h P(s)=\frac{1}{s}\left(1-\frac{\g}{s^{a+1}}+\cdots\right),
\eeq
and so
\beq
P(t)=1-\frac{\g\,t^{a+1}}{(a+1)!}+\cdots
\eeq
The departure of the survival probability from unity is slower
when the initial distance $a$ between the particle and the trap is larger,
as could be expected.

More interesting is the asymptotic decay of the survival probability
at late times, which is governed by the behavior of~(\ref{scres}) as $s\to0$.
We have
\beq
\h P(s)\approx\frac{b}{\sqrt{s}},
\eeq
and so
\beq
P(t)\approx\frac{b}{\sqrt{\pi t}},
\label{sasc}
\eeq
with
\beq
b=a+\frac{2}{\g}.
\label{bdef}
\eeq
The survival probability $P(t)$ exhibits the universal decay
in $1/\sqrt{t}$ of the persistence probability that a one-dimensional random walker
has not returned to its starting point up to time $t$.
The amplitude $b$ can be interpreted as an {\it effective distance}
between the trap and the site where the particle is launched,
which is just $a$ in the limit of a perfect trap ($\g\to\infty$),
but gets renormalized to a larger value whenever the trap is imperfect ($\g$ finite).

\subsection{A quantum particle}
\label{chainq}

Consider now a quantum particle propagating on the chain
in the presence of a single trap at the origin.
The strength $\g$ of the trap is now the amplitude of
an imaginary {\it optical potential} describing the absorbing power of the trap.
The tight-binding equation~(\ref{wqdif}) thus becomes non-Hermitian:
\beq
\ii\,\dif{\p_n(t)}{t}=\p_{n+1}(t)+\p_{n-1}(t)-\ii\g\delta_{n0}\,\p_0(t).
\label{qdif}
\eeq
Complex optical potentials were introduced long ago in quantum mechanics
in order to describe the inelastic scattering or the absorption of particles,
in analogy with complex refraction indices in optics.
In the present framework,
the effective description of the trap by an optical potential
can be derived by first representing the trap as a
zero-temperature phonon bath and then tracing out the degrees of freedom
of the bath~\cite{P2,RMP,Lindenberg,Huber,Silbey}.
A more formal approach is based on deriving a Lindblad dynamics
for the particle coupled to the bath~\cite{Selsto1,Selsto2,Diaz}.

We again assume that the particle is launched at site $a\ge0$ at time $t=0$.
We are again interested in the survival probability of the quantum particle
up to time $t$.
This quantity reads
\beq
\P(t)=\sum_n\abs{\p_n(t)}^2=1-2\g\!\int_0^t\abs{\p_0(t')}^2\d t'.
\label{qsint}
\eeq
The differential equation~(\ref{qdif})
can be solved in the very same way as its classical analogue~(\ref{cdif}).
We thus obtain
\beq
\h\p(q,s)=\frac{\ii(\e^{-\ii qa}-\g\,\h\p_0(s))}{\ii s-2\cos q}
\eeq
and
\beq
\h\p_0(s)=\frac{1}{\g+\sqrt{s^2+4}}
\left(\frac{\sqrt{s^2+4}-s}{2\ii}\right)^{\!a}.
\label{p0q}
\eeq

There is a noticeable qualitative difference between the quantum case under study
and its classical counterpart.
In the present situation,
the survival probability tends to a non-zero asymptotic value $\P_\infty$
in the limit of infinitely large times:
\beq
\P_\infty=1-2\g I,\qquad
I=\int_0^\infty\abs{\p_0(t)}^2\d t,
\label{qsinf}
\eeq
whose dependence on the trapping strength $\g$
and on the initial distance $a$ will now be worked out explicitly.
The integral $I$ can be evaluated directly in Laplace space,
by means of the following formula:
\beq
\int_0^\infty f(t)g(t)\,\e^{-\eps t}\,\d t=\int\frac{\d s}{2\pi\ii}\,\h f(s)\h g(\eps-s)
\quad(0<\Re s<\eps).
\label{ppiden}
\eeq
The above identity is a regularized form of the Parseval-Plancherel identity
for Fourier transforms, with $\eps>0$ acting as a regulator.
It holds as soon as
the Laplace transforms $\h f(s)$ and $\h g(s)$ are analytic in the half plane $\Re s>0$.
Setting $f(t)=\p_0(t)$ and $g(t)=\pbar_0(t)$ and taking the $\eps\to0$ limit,
so that $s=\ii y+0$ lies to the immediate right of the imaginary axis, we obtain
\beq
I=I_1+I_2(a),
\eeq
with
\beqa
{\hskip 14pt}I_1&=&\frac{1}{\pi}\int_0^2\frac{\d y}{(\g+\sqrt{4-y^2})^2},
\nonumber\\
I_2(a)&=&\frac{1}{\pi}\int_2^\infty\frac{\d y}{\g^2+y^2-4}
\left(\frac{y-\sqrt{y^2-4}}{2}\right)^{2a}.
\eeqa

In the limit where the initial distance gets large ($a\to\infty$),
$I_2(a)$ tends to zero, and so $I\to I_1$.

The explicit expressions of the above integrals depend
on the position of $\g$ with respect to 2.

\cas
$\g<2$.
Setting $w=\sqrt{4-\g^2}$, we obtain
\beqa
{\hskip 12pt}\pi I_1&=&\!-\frac{2}{w^2}+\frac{2}{w^3}\ln\frac{2+w}{2-w},
\nonumber\\
\pi I_2(0)&=&\frac{1}{2w}\ln\frac{2+w}{2-w},
\nonumber\\
\pi I_2(1)&=&\!-1+\frac{\pi\g}{4}
+\left(-\,\frac{1}{2w}+\frac{w}{4}\right)\ln\frac{2+w}{2-w},
\nonumber\\
\pi I_2(2)&=&\frac{8}{3}-w^2-\left(\frac{1}{2}-\frac{w^2}{4}\right)\pi\g
+\left(\frac{1}{2w}-w+\frac{w^3}{4}\right)\ln\frac{2+w}{2-w},
\nonumber\\
\pi I_2(3)&=&\!-\frac{21}{5}+\frac{14w^2}{3}-w^4
+\left(\frac{3}{4}-w^2+\frac{w^4}{4}\right)\pi\g
\nonumber\\
&+&\left(-\,\frac{1}{2w}+\frac{9w}{4}-\frac{3w^3}{2}
+\frac{w^5}{4}\right)\ln\frac{2+w}{2-w},
\label{ginf}
\eeqa
and so on.

In the weak-trapping regime ($\g\to0$),
the asymptotic survival probability departs from unity as
\beq
\P_\infty\approx1-\frac{2\g\abs{\ln\g}}{\pi}
\label{sweak}
\eeq
for any finite initial distance $a$ (including $a=0$),
whereas the limiting result for $a=\infty$, i.e., $I=I_1$,
yields a halved amplitude, i.e.,
\beq
\P_\infty\approx1-\frac{\g\abs{\ln\g}}{\pi}\qquad(a=\infty).
\label{sweakinf}
\eeq
In other words, the limits $a\to\infty$ and $\g\to0$ do not commute.

\cas
$\g>2$.
Setting $u=\sqrt{\g^2-4}$, we now have
\beqa
{\hskip 12pt}\pi I_1&=&\frac{2}{u^2}-\frac{4}{u^3}\arctan\frac{u}{2},
\nonumber\\
\pi I_2(0)&=&\frac{1}{u}\arctan\frac{u}{2},
\nonumber\\
\pi I_2(1)&=&\!-1+\frac{\pi\g}{4}
-\left(\frac{1}{u}+\frac{u}{2}\right)\arctan\frac{u}{2},
\nonumber\\
\pi I_2(2)&=&\frac{8}{3}+u^2
-\left(\frac{1}{2}+\frac{u^2}{4}\right)\pi\g
+\left(\frac{1}{u}+2u+\frac{u^3}{2}\right)\arctan\frac{u}{2},
\nonumber\\
\pi I_2(3)&=&\!-\frac{21}{5}-\frac{14u^2}{3}-u^4
+\left(\frac{3}{4}+u^2+\frac{u^4}{4}\right)\pi\g
\nonumber\\
&-&\left(\frac{1}{u}+\frac{9u}{2}+3u^3
+\frac{u^5}{2}\right)\arctan\frac{u}{2},
\label{gsup}
\eeqa
and so on.
These expressions are the analytic continuations of~(\ref{ginf}) as $w\to\pm\ii u$,
as it should be.

In the strong-trapping regime ($\g\to\infty$),
the behavior of the asymptotic survival probability depends
in a crucial way on the initial distance $a$.
If the particle is launched on the trap itself,
the survival probability decays to zero~as
\beq
\P_\infty\approx\frac{1}{2\g^2}\qquad(a=0).
\label{sstrongzero}
\eeq
If the particle is launched at any other site,
the survival probability goes to unity as
\beq
\P_\infty\approx1-\frac{16a^2}{\pi(4a^2-1)\g}\qquad(a>0).
\label{sstrong}
\eeq
We have in particular
\beq
\P_\infty\approx1-\frac{4}{\pi\g}
\label{sstronginf}
\eeq
in the limit of an infinitely large initial distance.

\begin{figure}[!ht]
\begin{center}
\includegraphics[angle=-90,width=.55\linewidth]{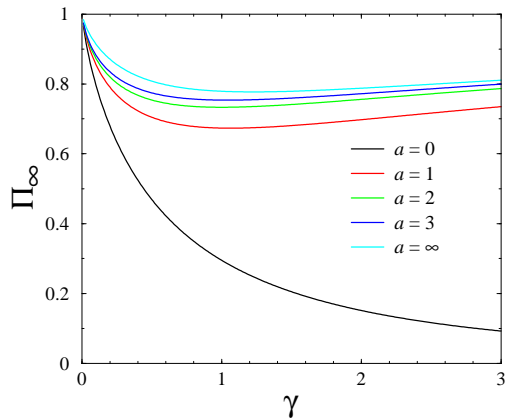}
\caption{\small
Asymptotic survival probability $\P_\infty$
of the quantum particle against trapping strength $\g$,
for several values of the initial distance $a$ between the particle and the trap.
Bottom to top: $a=0$, 1, 2, 3, and $\infty$
(Color online).}
\label{qsigma}
\end{center}
\end{figure}

Figure~\ref{qsigma} shows a plot of $\P_\infty$ against the trapping strength $\g$,
for several values of the initial distance $a$ between the particle and the trap.
This figure illustrates the most salient features of the asymptotic survival probability.
When the quantum particle is launched on the trap itself ($a=0$),~$\P_\infty$
decreases monotonically from unity (see~(\ref{sweak}))
to zero (see~(\ref{sstrongzero})).
As soon as the particle is launched at a different site ($a\ne0$),
$\P_\infty$ has a non-monotonic dependence on the trapping strength.
It starts decreasing from unity for small~$\g$ (see~(\ref{sweak})),
reaches a non-trivial minimum $\P_{\infty,\min}$
for some finite trapping strength $\g_\min$,
before it slowly returns to unity in the strong-trapping regime (see~(\ref{sstrong})).

The values of the asymptotic survival probability $\P_\infty$
in the borderline situation where $\g=2$ can be reached
by letting either $w\to0$ in~(\ref{ginf}) or $u\to0$ in~(\ref{gsup}).
The numbers thus obtained are listed in Table~\ref{qtable}
for initial distances $a=0$, 1, 2, 3, and $\infty$,
as well as the coordinates $\g_\min$ and $\P_{\infty,\min}$
of the minimum of the curve $\P_\infty(\g)$.
The quantities $\P_\infty(\g=2)$ and $\P_{\infty,\min}$
show a fast convergence in $1/a^2$ toward their values
in the $a=\infty$ limit.
For $a=2$, $\g_\min=1$ exactly, while
$\P_{\infty,\min}=1/2+2/\pi-5\ln(2+\sqrt3)/(3\pi\sqrt3)$.

\begin{table}[!ht]
\begin{center}
\begin{tabular}{|c||c||c|c|}
\hline
$a$&$\P_\infty(\g=2)$&$\g_\min$&$\P_{\infty,\min}$\\
\hline
0&\hfill$1-8/(3\pi)=0.15117$&$\infty$&$0$\\
\hline
1&\hfill$16/(3\pi)-1=0.69765$&1.06254&0.67338\\
\hline
2&\hfill$5-40/(3\pi)=0.75586$&1&0.73324\\
\hline
3&\hfill$272/(15\pi)-5=0.77201$&1.04211&0.75369\\
\hline
$\infty$&\hfill$1-2/(3\pi)=0.78779$&1.22574&0.77720\\
\hline
\end{tabular}
\end{center}
\caption{\small Values of the asymptotic survival probability $\P_\infty$
of the quantum particle in the borderline situation $\g=2$ (Column~2),
and coordinates $\g_\min$ and $\P_{\infty,\min}$ (Columns~3 and~4)
of the minimum of the curve $\P_\infty(\g)$,
for several initial distances $a=0$, 1, 2, 3, and $\infty$
between the particle and the trap.}
\label{qtable}
\end{table}

The most striking feature of the above results
is the non-monotonic dependence of the asymptotic survival probability $\P_\infty$
on the trapping strength~$\g$,
which eventually returns to unity in the strong-trapping regime ($\g\to\infty$).
The paradoxical inefficiency of a quantum trap in the nominal strong-trapping regime
can be understood in the following terms.
For large $\g$ and $a>0$, the wavefunction $\p_0(t)$ on the trap
can be expected to be proportional to $1/\g$,
and so the integrand in~(\ref{qsint}),~(\ref{qsinf}) is proportional to $1/\g^2$,
and the differences $1-\P(t)$ and $1-\P_\infty$
are proportional to $1/\g$.

In the present situation,
it can be checked explicitly that the above line of reasoning is correct.
To leading order as $\g\to\infty$,~(\ref{p0q}) indeed reads
\beq
\g\,\h\p_0(s)\approx\left(\frac{\sqrt{s^2+4}-s}{2\ii}\right)^{\!a},
\eeq
and yields
\beq
\g\,\p_0(t)\approx a\,\ii^{-a}\,\frac{J_a(2t)}{t},
\eeq
where $J_a$ is again the Bessel function.
Inserting the above estimate into~(\ref{qsinf}),
and using the integral~\cite[(6.574.2), p.~692]{gr}
\beq
\int_0^\infty\left(\frac{J_a(x)}{x}\right)^2\d x=\frac{4}{\pi(4a^2-1)},
\eeq
we recover~(\ref{sstrong}).

\section{A finite ring with a single trap}
\label{ring}

In this section we investigate the motion of a classical and a quantum particle
on a finite ring with a single trap.

\subsection{A classical particle}
\label{ringc}

Consider a classical particle diffusing
on a ring of $N$ sites ($n=0,\dots,N-1$),
with periodic boundary conditions ($N\equiv0$).
In the presence of a single trap at the origin,
the probability $p_n(t)$ for the particle to be at site $n$ at time $t$
still obeys~(\ref{cdif}).
It will be sufficient for our purpose to
look for solutions decaying as $\exp(-Et)$,
i.e., for the eigenmodes of the stationary equation
\beq
-E\,p_n=p_{n+1}+p_{n-1}-2p_n-\g\delta_{n0}\,p_0,
\label{cstat}
\eeq
with periodic boundary conditions.
Setting
\beq
E=2(1-\cos q),
\label{cdisp}
\eeq
the general solution of~(\ref{cstat}) reads
\beq
p_n=A\,\e^{\ii nq}+B\,\e^{-\ii nq}
\label{gsol}
\eeq
for $n=0,\dots,N$ with $N\equiv0$.
Boundary conditions yield
\beqa
{\hskip 13pt}A+B&=&A\,\e^{\ii Nq}+B\,\e^{-\ii Nq},
\nonumber\\
\g(A+B)&=&A\,\e^{-\ii q}(\e^{\ii Nq}-1)+B\,\e^{\ii q}(\e^{-\ii Nq}-1).
\label{cbcds}
\eeqa

Eigenmodes have a definite parity,
i.e., they are either even $\even$ or odd~$\odd$
in the exchange $n\leftrightarrow N-n$.

\cas In the even sector, we have $p_{N-n}=p_n$, i.e., $B=A\,\e^{\ii Nq}$.
The amplitude $p_0=A+B$ on the trap is non-vanishing.
The second line of~(\ref{cbcds}) yields the quantization condition
\beq
\e^{\ii Nq}=\frac{2\sin q+\ii\g}{2\sin q-\ii\g},\qquad\hbox{i.e.,}
\;\;\g=2\sin q\,\tan\frac{Nq}{2}.
\label{ceven}
\eeq

\cas In the odd sector, we have $p_{N-n}=-p_n$, i.e., $B=-A\,\e^{\ii Nq}$.
The amplitude on the trap vanishes.
This yields the quantization condition
\beq
\e^{\ii Nq}=1,
\label{codd}
\eeq
irrespective of $\g$.
We thus obtain the explicit momenta $q_k^\odd=2\pi k/N$, i.e.,
\beq
E_k^\odd=2\left(1-\cos\frac{2\pi k}{N}\right).
\eeq

All eigenvalues are positive,
and the even and odd spectra are intertwining:

\cas If $N$ is odd, say $N=2M+1$, there are $M$ odd and $M+1$ even modes:
\beq
0<E_1^\even<E_1^\odd<\cdots<E_M^\even<E_M^\odd<4<E_{M+1}^\even.
\eeq

\cas If $N$ is even, say $N=2M+2$, there are $M$ odd and $M+2$ even modes:
\beq
0<E_1^\even<E_1^\odd<\cdots<E_M^\even<E_M^\odd<E_{M+1}^\even<4<E_{M+2}^\even.
\eeq
In both cases the largest eigenvalue in the even sector is larger
than the band edge ($E_\max=4$).
It therefore corresponds to a purely imaginary momentum.

For a generic initial condition,
the survival probability decays exponentially fast, as
\beq
P(t)\sim\e^{-\lambda t},\qquad\lambda=E_1^\even.
\eeq

When the ring size $N$ is large,
we have $q_1^\even\approx\pi/N$, and so the decay rate~$\lambda$ scales as
\beq
\lambda\approx\frac{\pi^2}{N^2},
\label{cdir}
\eeq
irrespective of $\g$.

The decay rate however exhibits a non-trivial scaling in the regime
where~$N$ is large, while the trapping strength $\g$ is small,
so that the product
\beq
X=N\g
\label{Xdef}
\eeq
is fixed.
In this regime, we have $q_1^\even\approx\theta/N$,
and so the decay rate obeys the scaling law
\beq
\lambda\approx\frac{\theta^2}{N^2},
\label{csca}
\eeq
where $\theta$ is an implicit function of $X$ given by (see~(\ref{ceven}))
\beq
2\theta\,\tan\frac{\theta}{2}=X.
\eeq

\cas
For $X\ll1$, we have $\theta^2=X-X^2/12+\cdots$, and so
\beq
\lambda\approx\frac{\g}{N}\left(1-\frac{N\g}{12}+\cdots\right)\qquad(N\g\ll1).
\eeq

\cas
For $X\gg1$, we have $\theta=\pi-4\pi/X+\cdots$, and so
\beq
\lambda\approx\frac{\pi^2}{N^2}\left(1-\frac{8}{N\g}+\cdots\right)\qquad(N\g\gg1),
\eeq
in agreement with~(\ref{cdir}) to leading order.
The scaling function $\theta(X)$ will be plotted in Figure~\ref{rates},
together with its quantum analogue (see~(\ref{qsca})).

\subsection{A quantum particle}
\label{ringq}

Consider now a quantum particle on a finite ring of $N$ sites,
in the presence of a single trap at the origin.
This problem has already been tackled in several earlier works~\cite{RMP,AMB,GAS}.

For the present purpose, it will be sufficient to
look for harmonic solutions to~(\ref{qdif}),
proportional to $\exp(-\ii\eps t)$,
i.e., for the eigenmodes of the stationary equation
\beq
\eps\,\p_n=\p_{n+1}+\p_{n-1}-\ii\g\delta_{n0}\,\p_0,
\label{qstat}
\eeq
with periodic boundary conditions.
The general solution of~(\ref{qstat}) is still given by~(\ref{gsol}),
albeit with the dispersion law~(\ref{qdisp}).

\cas In the even sector, the quantization condition now reads
\beq
\e^{\ii Nq}=\frac{2\sin q-\g}{2\sin q+\g},\qquad\hbox{i.e.,}
\;\;\g=-2\ii\sin q\,\tan\frac{Nq}{2}.
\label{qeven}
\eeq

\cas In the odd sector, the quantization~(\ref{codd}) still holds,
irrespective of the trapping strength $\g$.
We thus obtain the real energy eigenvalues
\beq
\eps_k^\odd=2\cos\frac{2\pi k}{N}.
\eeq

\begin{figure}[!ht]
\begin{center}
\includegraphics[angle=-90,width=.55\linewidth]{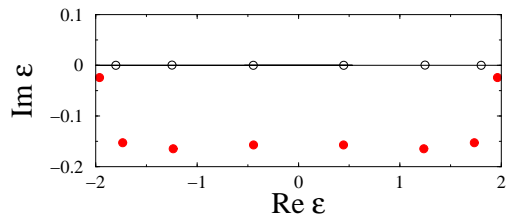}
\caption{\small
Energy spectrum of a ring of $N=14$ sites with $\g=1$
in the complex~$\eps$-plane.
The scale on the imaginary axis is expanded 5 times.
Empty black symbols: 6 real eigenvalues of the odd sector.
Full red symbols: 8 complex eigenvalues of the even sector
(Color online).}
\label{ring14}
\end{center}
\end{figure}

The main qualitative difference between the quantum case
and its classical counterpart
is that the energy eigenvalues in the even sector are now
complex numbers with negative imaginary parts,
while those in the odd sector remain real.
Figure~\ref{ring14} shows a plot of the energy spectrum
for a ring of $N=14$ sites with $\g=1$.

Consider an arbitrary initial condition,
defined by the normalized wavefunction $\p_n(0)=\phi_n$,
and introduce the projections
\beq
\phi_n^\eo=\frac{1}{2}(\phi_n\pm\phi_{N-n})
\eeq
of the initial wavefunction onto the even and odd sectors.

The projection $\phi_n^\odd$ undergoes a unitary evolution,
i.e., it is the superposition
of harmonic components in $\exp(-\ii\eps_k^\odd t)$,
where the energy eigenvalues $\eps_k^\odd$ are real.
The projection $\phi_n^\even$ undergoes a non-unitary evolution,
i.e., it is the superposition
of decaying components in $\exp(-\ii\eps_k^\odd t)$,
where the energy eigenvalues $\eps_k^\odd$ are complex numbers
with negative imaginary parts.
The survival probability $\P(t)$ therefore
tends to a non-zero asymptotic value~$\P_\infty$,
which is nothing but the probability for the particle to be initially in the odd sector:
\beq
\P_\infty=\sum_{n=1}^{N-1}\abs{\phi_n^\odd}^2.
\label{sinf}
\eeq

In order to make the connection with the infinite chain
(see Section~\ref{chain}),
let us assume that the particle is launched at some site $a$ on the ring.
We have then generically
\beq
\P_\infty=\frac{1}{2}.
\label{sring}
\eeq
There are two exceptions to this rule.
If $a=0$ (the particle is launched on the trap itself),
$\phi_n^\odd$ vanishes and we have $\P_\infty=0$.
More surprisingly, if $N$ is even and $a=N/2$
(the particle is launched exactly at the antipode of the trap),
$\phi_n^\odd$ also vanishes and we again have $\P_\infty=0$.
This is a mere interference phenomenon in a discrete quantum system.
It is also worth noticing that~(\ref{sring})
is smaller than the corresponding result on the infinite chain (see~Table~\ref{qtable}),
where the minimum $\P_{\infty,\min}$ depends on the distance $a$,
but is always larger than $1/2$.
This difference can be attributed to the fact that the particle
may escape to infinity on the infinite chain,
whereas it is bound to make repeated returns to the trap in the ring geometry.

The excess survival probability $\P(t)-\P_\infty$
decays exponentially fast, as
\beq
\P(t)-\P_\infty\sim\e^{-\lambda t},
\eeq
where the decay rate $\lambda$ is dictated by the eigenvalue in the even sector
whose imaginary part is the smallest, i.e.,
\beq
\lambda=-2\Im\eps_1^\even.
\eeq

The decay rate $\lambda$ again exhibits a non-trivial scaling
if the ring size~$N$ is large,
while the trapping strength $\g$ is small,
so that their product $X$ is kept constant (see~(\ref{Xdef})).
We have $q_1^\even\approx\zeta/N$, and so the decay rate obeys the scaling law
\beq
\lambda\approx\frac{2\Im\zeta^2}{N^2}.
\label{qsca}
\eeq
Using~(\ref{qeven}), one finds that $\zeta$ is an implicit complex function of $X$ given by
\beq
2\zeta\,\tan\frac{\zeta}{2}=\ii X.
\eeq

\cas
For $X\ll1$, we have $\zeta^2=\ii X+X^2/12-\ii X^3/180+\cdots$, and so
\beq
\lambda\approx\frac{2\g}{N}\left(1-\frac{(N\g)^2}{180}+\cdots\right)
\qquad(N\g\ll1).
\eeq

\cas
For $X\gg1$, we have $\zeta=\pi(1+4\ii/X-16/X^2-16(12-\pi^2)\ii/(3X^3)+\cdots)$, and so
\beq
\lambda\approx\frac{16\pi^2}{N^3\g}
\left(1-\frac{4(24-\pi^2)}{3(N\g)^2}+\cdots\right)\qquad(N\g\gg1).
\eeq

The decay rate $\lambda$ thus shows a non-monotonic dependence
on the trapping strength.
It starts increasing linearly as $\g/N$ for small $\g$,
reaches a maximum for $X_\max=N\g_\max=9.2235$, where $N^2\lambda_\max=11.0846$,
and then falls off as $1/(N^3\g)$.
The power-law decay
\beq
\lambda\approx\frac{16\pi^2}{N^3\g}
\label{qlring}
\eeq
holds more generally for $N$ large and any finite value of $\g$.

\begin{figure}[!ht]
\begin{center}
\includegraphics[angle=-90,width=.55\linewidth]{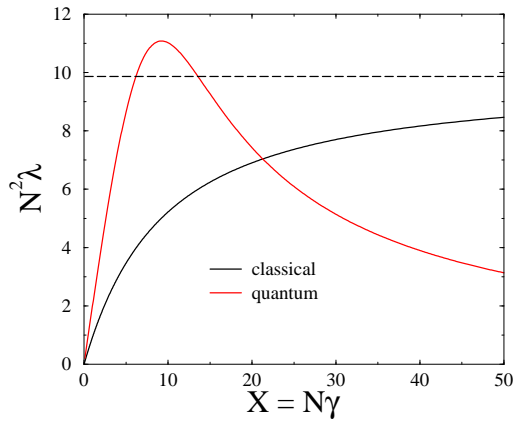}
\caption{\small
Rescaled decay rate $N^2\lambda$ against rescaled trapping strength $X=N\g$
for a classical and a quantum particle on a ring with a single trap.
The corresponding scaling functions are
$\theta^2$ for the classical case (monotonic, see~(\ref{csca}))
and $2\Im\zeta^2$ for the quantum one (non-monotonic, see~(\ref{qsca})).
Dashed line: asymptote at $\pi^2$ for the classical case
(Color online).}
\label{rates}
\end{center}
\end{figure}

Figure~\ref{rates} shows a comparison between the scaling forms
of the decay rate~$\lambda$ in the classical case (see~(\ref{csca}))
and in the quantum one (see~(\ref{qsca})).

\section{A finite ring with several traps}
\label{rings}

We have seen in Section~\ref{ringq} that a quantum particle
on a ring with a single trap
survives forever with a non-zero probability $\P_\infty$.
This is due to the existence of {\it avoiding modes}
which are insensitive to the presence of the trap,
because their amplitude on the trap vanishes.
Stated otherwise, avoiding modes are solutions of the stationary equation~(\ref{qstat})
whose energy eigenvalue $\eps$ is {\it real}.
This phenomenon has no classical analogue:
on a finite sample, a single trap always causes the survival probability
of a classical walker to decay exponentially fast to zero.

For a single trap, avoiding modes are easy to characterize:
these are the modes pertaining to the odd sector.
Remarkably enough, avoiding modes can exist in the presence of more than one trap.
Consider the example of a ring of~$N$ sites, in the case where $N=2M$ is even,
with two traps at the antipodal positions $n=0$ and $n=M$.
All the modes pertaining to the odd sector
are still avoiding modes in the presence of these two traps.

In general, the asymptotic survival probability $\P_\infty$ of the quantum particle
is given by the square norm of the projection
of its initial state $\ket{\phi}$ onto the linear subspace of avoiding modes.
In other words, the result~(\ref{sinf}) generalizes~to
\beq
\P_\infty=\sum_{\ket{e}}\abs{\braket{e}{\phi}}^2,
\eeq
where $\ket{e}$ runs over a basis of avoiding modes.

The existence of avoiding modes in the presence of several traps on a ring
has already been demonstrated on examples~\cite{RMP,AMB}.
Our goal is to treat the combinatorics of these modes in a systematic way.
Consider the $2^N$ trap configurations on the ring.
Each configuration is encoded in a sequence of occupation numbers $\eta_n\in\{0,1\}$,
with $\eta_n=1$ meaning there is a trap at site $n$
and $\eta_n=0$ there is no trap at site $n$.
We assume all traps have the same strength $\g$.
For each trap configuration, we consider the stationary tight-binding equation
\beq
\eps\,\p_n=\p_{n+1}+\p_{n-1}-\ii\g\eta_n\,\p_n,
\label{tbsys}
\eeq
with periodic boundary conditions ($N\equiv0$),
and we address the following question:
{\it Does~(\ref{tbsys}) have avoiding modes?}

If there is no trap, all the $N$ eigenmodes of~(\ref{tbsys}) are avoiding.
Let us henceforth assume there is at least one trap, and proceed in two steps:

\cas {\it At the local level},
consider a cluster of consecutive sites without traps.
Define the cluster length $L\ge2$ as the number of those sites plus one,
and renumber the sites as $n=1,\dots,L-1$.
This cluster supports avoiding modes of the form
\beq
\psi_n=\sin qn,
\eeq
with $q=k\pi/L$ with $k=1,\dots,L-1$.

\cas {\it At the global level},
for a given trap configuration,
define $\ell$ as the greatest common divisor (GCD)
of all the cluster lengths $L_1, L_2,\dots$
The condition for the configuration to support at least one avoiding mode
simply reads $\ell\ge2$.
In the marginal situation where $\ell=2$,
one must impose the extra condition that $N$ is a multiple of 4.

For a ring of $N$ sites, we are thus led to define the following combinatorial numbers:

\cas The number $m(N,\ell)$ of non-empty trap configurations
such that $\ell$ is the GCD of all the cluster lengths.

\cas The number $M(N)$ of trap configurations supporting avoiding modes.

These numbers can be calculated recursively as follows.
First, $m(N,\ell)$ is non-zero only if $\ell$ is a divisor of $N$, and we have
\beq
m(N,\ell)=\ell\;\nu(N/\ell),
\label{combi1}
\eeq
where $\nu(N)$ is a short-hand for $m(N,1)$.
Then, any non-empty configuration corresponds a unique $\ell$,
hence the sum rule\footnote{The notation $\ell\mid N$
means that $\ell$ is a divisor of $N$, i.e., that $N$ is a multiple of $\ell$.}
\beq
2^N=1+\sum_{\ell\mid N}m(N,\ell).
\label{combi2}
\eeq
Singling out the term with $\ell=1$, we obtain the following recursion relation:
\beq
\nu(N)=2^N-1-{\sum_{\ell\mid N}}'\,\ell\,\nu(N/\ell),
\eeq
where the sum runs over all the divisors $\ell$ of $N$, except $\ell=1$.
The above relation determines all the $\nu(N)$ from the initial value $\nu(1)=1$.

The number of configurations supporting avoiding modes is
\beq
M(N)=1+{\sum_{\ell\mid N}}'\,\ell\;\nu(N/\ell),
\eeq
where the sum runs over the divisors $\ell$ of $N$ except $\ell=1$,
and $\ell=2$ if $N$ is not a multiple of 4.
The above expression simplifies to
\beq
M(N)=\left\{
\matrix{
2^N-\nu(N)-2\nu(N/2)\quad&(\hbox{$N$ even and $N/2$ odd}),\cr
2^N-\nu(N)\hfill&(\hbox{else}).\hfill
}\right.
\label{mnres}
\eeq

\begin{figure}[!ht]
\begin{center}
\includegraphics[angle=-90,width=.55\linewidth]{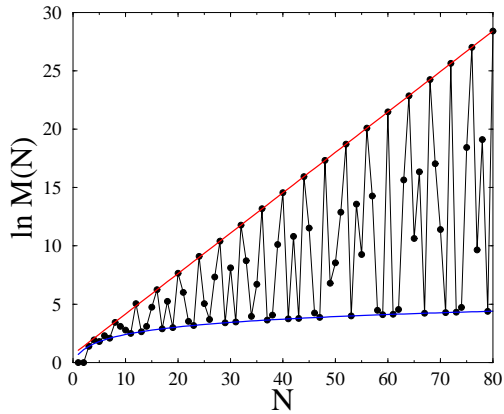}
\caption{\small
Logarithmic plot of the number $M(N)$ of trap configurations
supporting avoiding modes on a ring of $N$ sites, against $N$.
Symbols: exact numbers (see~(\ref{mnres})).
Lower blue curve: minimal values (see~(\ref{mmin})),
reached when $N$ is a prime number.
Upper red line: maximal values (see~(\ref{mmax})),
reached exactly when $N$ is a power of 2,
and approximately whenever $N$ is a multiple of 4
(Color online).}
\label{mn}
\end{center}
\end{figure}

The erratic dependence of $M(N)$ on the ring size $N$,
illustrated in Table~\ref{mntable} and Figure~\ref{mn},
is reminiscent of the behavior of the classical number-theoretic functions.
In particular:

\begin{table}[!ht]
\begin{center}
\begin{tabular}{|c|c||c|c||c|c||c|c|}
\hline
$N$&$M(N)$ & $N$&$M(N)$ & $N$&$M(N)$ & $N$&$M(N)$\\
\hline
1&1 & 6&10 & 11&12 & 16&511\\
2&1 & 7&8 & 12&154 & 17&18\\
3&4 & 8&31 & 13&14 & 18&190\\
4&7 & 9&22 & 14&22 & 19&20\\
5&6 & 10&16 & 15&114 & 20&2092\\
\hline
\end{tabular}
\end{center}
\caption{\small Numbers $M(N)$
of trap configurations supporting avoiding modes for rings with up to $N=20$ sites.}
\label{mntable}
\end{table}

\cas If $N$ is an odd prime number, only the empty configuration
and the $N$ configurations with a single trap ($\ell=N$)
support avoiding modes.
We thus obtain the linear behavior
\beq
M(N)=N+1.
\label{mmin}
\eeq

\cas If $N$ is a power of 2 (albeit not 2 itself),
we have the exponential behavior
\beq
M(N)=2^{1+N/2}-1.
\label{mmax}
\eeq
This number of configurations grows as the square root
of the total number~$2^N$ of trap configurations.
The above growth law is approximately valid whenever~$N$ is a multiple of 4.

\section{The infinite chain with a fixed density of traps}
\label{1d}

We now consider the physically more realistic situation of a classical
or a quantum particle propagating on the infinite chain,
with a random distribution of traps with given density.
The higher-dimensional situation will be dealt with in Section~\ref{highd}.

\subsection{A classical particle}
\label{1dc}

We start with a reminder on the case of a classical particle.
We assume that all traps have the same strength $\g$,
so that we are facing a dilution disorder,
characterized by the trap density $c$,
and encoded in the random occupation numbers:
\beq
\eta_n
=\left\{\matrix{
1&\hbox{with prob.}\ c,\hfill\cr
0&\hbox{with prob.}\ 1-c,}\right.
\label{binary}
\eeq
with $\eta_n=1$ meaning there is a trap at site $n$,
and $\eta_n=0$ there is no trap at site $n$.
We are led to consider the stationary equation
\beq
-E\,p_n=p_{n+1}+p_{n-1}-2p_n-\g\eta_n\,p_n.
\label{c1}
\eeq

On any finite sample,
the survival probability decays exponentially fast, as
\beq
P(t)\sim\e^{-\lambda t},\qquad\lambda=E_1,
\eeq
where $E_1$ is the smallest eigenvalue of~(\ref{c1}).

It is known from the theory of Lifshitz tails~\cite{lif,lgp,fp}
that the smallest eigenvalue~$E_1$ corresponds to the lowest mode on the largest
{\it Lifshitz region}, i.e., the largest interval which is free of traps.
Let us focus on such a region of $N+1$ consecutive sites,
and renumber them as $n=0,\dots,N$.
We have therefore $\eta_n=0$ for $n=0,\dots,N$,
whereas the configuration on the rest of the sample is left unspecified.
The solution of~(\ref{c1}) on the Lifshitz region reads (see~(\ref{gsol}))
\beq
p_n=A\,\e^{\ii nq}+B\,\e^{-\ii nq}\qquad(n=0,\dots,N),
\eeq
where the wavevector $q$ is related to $E$ by (see~(\ref{cdisp}))
\beq
E=2(1-\cos q).
\eeq
Along the lines of~\cite{NL}, let us parametrize
the boundary conditions imposed by the rest of the sample as
\beq
p_{-1}=Y_L p_0,\qquad p_{N+1}=Y_R p_N.
\label{cbcs}
\eeq
We thus obtain the quantization condition
\beq
\sin(N+2)q-(Y_L+Y_R)\sin(N+1)q+Y_LY_R\sin Nq=0.
\eeq
The wavevector corresponding to the lowest mode therefore reads
\beq
q_1=\frac{\pi}{N}\left(1+\frac{\alpha}{N}+\cdots\right),
\label{q1}
\eeq
with
\beq
\alpha=\frac{1}{Y_L-1}+\frac{1}{Y_R-1}.
\label{alpha}
\eeq
To leading order as the size $N$ of the Lifshitz region is large,
we therefore have the estimate
\beq
\lambda\approx\frac{\pi^2}{N^2}.
\label{clif}
\eeq
This result is formally identical to~(\ref{cdir}),
which applies to a ring with a single trap.
It holds irrespectively of the rest of the sample,
which only enters the amplitude $\alpha$ of the correction term
through the left and right boundary parameters $Y_L$ and~$Y_R$.

In the limit of an infinite system,
the behavior of the integrated density of states
\beq
H(E)=\int_0^E\rho(E')\,\d E'
\eeq
at low energy ($E\to0$) can be estimated a follows:
the lowest energy level $E_1$ is given by~(\ref{clif})
with an exponentially small probability of order $(1-c)^N$ per unit length.
We thus obtain the well-known Lifshitz tail~\cite{lif,lgp,fp}
\beq
H(E)\sim\exp\left(-\,\frac{\pi\abs{\ln(1-c)}}{\sqrt{E}}\right).
\label{lif1}
\eeq
Despite the rather heuristic nature of the above reasoning,
the asymptotic result~(\ref{lif1}) is fully correct.
Indeed it only misses an oscillatory prefactor of order unity.
This prefactor has been shown to be a periodic function
of $\pi/\sqrt{E}$ with unit period,
which reflects the discrete nature of the underlying chain,
and can be expressed in terms of the distribution
of the boundary parameters $Y_L$ and $Y_R$ at $E=0$~\cite{NL}.

The asymptotic decay of the survival probability $P(t)$
can be estimated by superposing the contributions
of Lifshitz regions of all sizes $N$,
weighted by their respective probabilities $(1-c)^N$.
Working with exponential accuracy and going to a continuum approximation, we get
\beq
P(t)\sim\int_0^\infty\exp\left(-\,\frac{\pi^2t}{N^2}-\abs{\ln(1-c)}\!N\right)\d N.
\eeq
Finally, evaluating the integral by the saddle-point method, we obtain
\beq
P(t)\sim\exp\left(-\,\frac{3}{2}\left(2\pi^2\abs{\ln(1-c)}^2t\right)^{1/3}\right).
\eeq
This is precisely the stretched exponential law of the survival probability announced
in~(\ref{pc}) for $d=1$, with
\beq
A_1=\frac{3}{2}(2\pi^2)^{1/3}\abs{\ln(1-c)}^{2/3}.
\label{a1res}
\eeq

\subsection{A quantum particle}
\label{1dq}

Turning to the case of a quantum particle,
we consider the stationary non-Hermitian tight-binding equation
\beq
\eps\,\p_n=\p_{n+1}+\p_{n-1}-\ii\g\eta_n\,\p_n,
\label{tb}
\eeq
where the disordered configuration of traps
is given by the binary distribution~(\ref{binary}).

Consider for a while a very long but finite sample.
The probability that it supports avoiding modes
decays exponentially with the sample length.
Neglecting these very improbable events,
the survival probability decays exponentially fast, as
\beq
\P(t)\sim\e^{-\lambda t},
\eeq
where the decay rate $\lambda$ is dictated by the eigenvalue
whose imaginary part is the smallest, i.e.,
\beq
\lambda=-2\Im\eps_1.
\eeq
Furthermore, $\eps_1$ is again expected to correspond to the lowest mode
on the largest Lifshitz region.

Let us therefore consider,
along the lines of Section~\ref{1dc},
a trap-free Lifshitz region of size $N$
(i.e., made of $N+1$ consecutive sites).
The wavevector $q_1$ corresponding to the lowest mode
on that region is still given by~(\ref{q1}),~(\ref{alpha})
in terms of the left and right boundary parameters $Y_L$ and $Y_R$.
We therefore obtain the estimate
\beq
\lambda\approx2\Im q_1^2\approx\frac{4\pi^2}{N^3}\,\Im\alpha,
\label{ql}
\eeq
with (see~(\ref{alpha}))
\beq
\alpha=\frac{1}{Y_L-1}+\frac{1}{Y_R-1}.
\label{qalpha}
\eeq

\begin{figure}[!ht]
\begin{center}
\includegraphics[angle=-90,width=.7\linewidth]{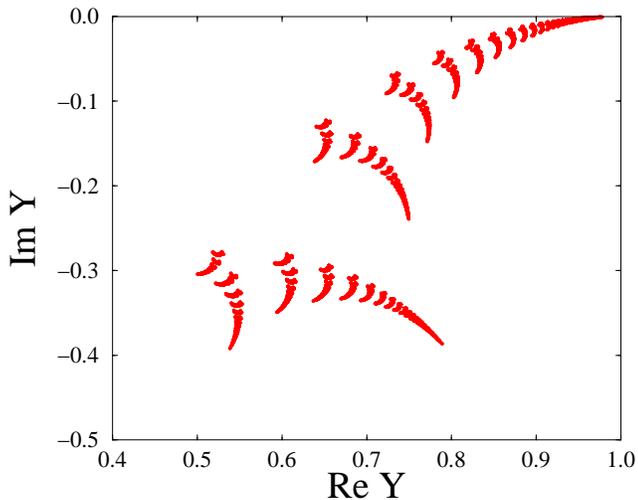}
\caption{\small
Fractal support of the invariant distribution of the Riccati variables $Y_n$
in the complex $Y$-plane for $\g=1/2$
(Color online).}
\label{yfractal}
\end{center}
\end{figure}

The result~(\ref{ql}) has two qualitative differences with respect to
its classical counterpart~(\ref{clif}).
First, the decay rate scales as $1/N^2$ in the classical situation,
and as $1/N^3$ in the quantum one.
These scaling laws already hold for a ring with a single trap
(see~(\ref{cdir}) and~(\ref{qlring})).
Second, the quantum decay rate~(\ref{ql}) is now a {\it fluctuating} quantity.
Indeed it depends {\it to leading order} on the rest of the sample
through the left and right boundary parameters $Y_L$ and~$Y_R$.
In order to proceed,
it is therefore necessary to determine the distribution of these boundary parameters.
This can be done as follows, along the lines of~\cite{NL}.
In the theory of one-dimensional disordered systems (see e.g.~\cite{JMLivre}),
it is customary to introduce Riccati variables,
defined as ratios of the wavefunction amplitudes at two consecutive sites:
\beq
Y_n=\frac{\psi_n}{\psi_{n+1}}.
\eeq
Right at the band edge ($q=0$, i.e., $\eps=2$),
these variables obey the recursion
\beq
Y_n=\frac{1}{2+\ii\g\eta_n-Y_{n-1}}
=\left\{\matrix{
\frad{1}{2+\ii\g-Y_{n-1}}&\hbox{with prob.}\ c,\hfill\cr
\frad{1}{2-Y_{n-1}}\hfill&\hbox{with prob.}\ 1-c.}\right.
\label{yrec}
\eeq
The Riccati variable $Y_n$ is therefore a function of the initial value $Y_0$
and of the random occupation numbers $\eta_1,\dots,\eta_n$.
In the $n\to\infty$ limit,
$Y_n$ has a non-trivial limit distribution in the complex $Y$-plane,
which is invariant under the random transformation~(\ref{yrec}).
Invariant distributions with complex support have already been met
in the realm of one-dimensional disordered systems
(see e.g.~\cite{HOZ}).
In the present situation of a binary dilution disorder,
the invariant distribution has a fractal support
which only depends on the trapping strength~$\g$,
and not on the trap density $c$.
This fractal set is shown in Figure~\ref{yfractal}
for a typical trapping strength ($\g=1/2$).

For the time being, let us keep considering a Lifshitz region
of fixed (large) size $N$.
If this region is embedded into an infinitely large disordered system,
the left and right boundary parameters $Y_L$ and $Y_R$ are independent of each other,
and each of them is distributed according to the above invariant distribution
in the complex $Y$-plane.
The distribution of the complex parameter $\alpha$ defined in~(\ref{qalpha})
is thus determined, at least in principle.
The smallest possible value~$\lambda_\min$ of the decay rate
is therefore given by the exact formula
\beq
\lambda_\min\approx\frac{8\pi^2 f}{N^3\g},
\label{lmin}
\eeq
where the factor $f$ reads
\beq
f=\min\,\Re F,
\label{fres}
\eeq
while the complex variables $F_n$ are related to the Riccati variables $Y_n$ through
\beq
F_n=\frac{\ii\g}{1-Y_n},
\eeq
and $\min\,\Re F$ has to be understood as being the minimum
of the real part of the support
of the invariant distribution of the variables $F_n$.
This support, to be shown in Figure~\ref{ffractal} below,
is nothing but the image in the complex $F$-plane
of the fractal set shown in Figure~\ref{yfractal}.

The expression~(\ref{lmin}) has the same leading scaling in $1/(N^3\g)$
as the result~(\ref{qlring}) for a ring with a single trap.
The factor $f$ only depends on the trapping strength $\g$,
and it is normalized so that it goes to unity in the $\g\to\infty$ limit.
In particular, $f$ is independent of the trap density $c$.

In order to evaluate the factor $f$ explicitly,
let us recast the recursion~(\ref{yrec})
in terms of the variables $F_n$ themselves:
\beq
F_n=\ii\g+\frac{F_{n-1}}{1+\eta_nF_{n-1}}
=\left\{\matrix{
\ii\g+\frad{F_{n-1}}{1+F_{n-1}}&\hbox{with prob.}\ c,\hfill\cr
\ii\g+F_{n-1}\hfill&\hbox{with prob.}\ 1-c.}\right.
\label{frec}
\eeq
The second line of the transformation~(\ref{frec})
is a translation by the constant imaginary quantity $\ii\g$.
As a consequence, the fractal support of the invariant distribution of the $F_n$
extends up to infinity in the imaginary direction of the complex $F$-plane,
and it consists on an infinite periodic array of identical components.
Figure~\ref{ffractal} shows the lower part
of this fractal support in the $F$-plane,
featuring the first component of the above mentioned periodic array,
for $\g=1/2$.

\begin{figure}[!ht]
\begin{center}
\includegraphics[angle=-90,width=.7\linewidth]{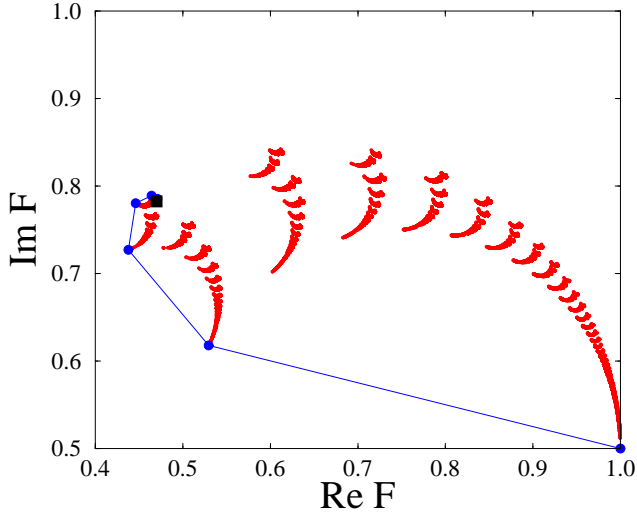}
\caption{\small
Lower part of the fractal support
of the invariant distribution of the variables $F_n$
in the complex $F$-plane for $\g=1/2$.
Blue symbols connected by a line: sequence of extremal points $F_k$
($k=1,2,\dots$).
Square: fixed point $F_\star$
(Color online).}
\label{ffractal}
\end{center}
\end{figure}

The factor $f$,
equal to the minimum of the real part of the fractal support,
can be determined by means of the following construction.
The lower right endpoint of the support is $F_1=1+\ii\g$.
This point is obtained by acting
on the point at infinity with the first line of~(\ref{frec}).
More generally, we are led to consider a discrete sequence of extremal points $F_k$
($k=1,2,\dots$),
shown in Figure~\ref{ffractal} as blue symbols connected by a line,
defined by iterating the first line of~(\ref{frec}),~i.e.,
\beq
F_k=\ii\g+\frac{F_{k-1}}{1+F_{k-1}},
\eeq
with $F_0=\infty$.
We thus obtain
\beqa
F_1&=&1+\ii\g,\quad F_2=\frac{\g^2+2+\ii\g(\g^2+5)}{\g^2+4},
\nonumber\\
F_3&=&\frac{\g^4+8\g^2+3+\ii\g(\g^4+11\g^2+14)}{(\g^2+1)(\g^2+9)},
\eeqa
and so on.
These points spiral around the stable fixed point
\beq
F_\star=\frac{\ii}{2}\left(\g+\sqrt{\g(\g-4\ii)}\right),
\label{Fstar}
\eeq
shown as a square in Figure~\ref{ffractal}, whose real part reads
\beq
f_\star=\sqrt{\frac{2\g}{\g+\sqrt{\g^2+16}}}.
\label{fstar}
\eeq

Let us give for further use the following explicit parametrization of the extremal points.
Introducing the complex variable $z$ through
\beq
\sqrt{\ii\g}=2\sinh z\quad(0<\Im z<\pi/2),
\label{defth}
\eeq
we have
\beq
F_k=\frac{2\sinh z\,\cosh(2k+1)z}{\sinh 2kz},
\label{fth}
\eeq
and especially
\beq
F_\star=\e^{2z}-1.
\label{fsth}
\eeq

It should now be clear from the construction shown in Figure~\ref{ffractal}
that the factor $f$ is the smallest of the real parts of the extremal points $F_k$.
In other words, we have
\beq
f=\Re F_{k(\g)},
\eeq
where the index $k(\g)$ of the point with smallest real part depends
on the trapping strength $\g$.

\cas $\g>1$.
In this first region, $F_2$ has the smallest real part, i.e., $k(\g)=2$.
We have therefore
\beq
f=\Re F_2=\frac{\g^2+2}{\g^2+4}\quad(\g>1).
\eeq
For large $\g$, this quantity departs from unity as
\beq
f=1-\frac{2}{\g^2}+\frac{8}{\g^4}+\cdots,
\label{fstrong}
\eeq
while the real part of the stable fixed point departs from unity as
\beq
f_\star=1-\frac{2}{\g^2}+\frac{14}{\g^4}+\cdots
\label{fsstrong}
\eeq

\cas $1/\sqrt{5}<\g<1$.
In this second region, $F_3$ has the smallest real part, i.e., $k(\g)=3$.
We have therefore
\beq
f=\Re F_3=\frac{\g^4+8\g^2+3}{(\g^2+1)(\g^2+9)},
\eeq
and so on.

\cas $\g\to0$.
In this regime of a weak trapping strength,
the extremal points come close to each other,
so that it is legitimate to use a continuum approach.
The variable $z$ becomes small in this regime, scaling as
\beq
z\approx\frac{\sqrt{\ii\g}}{2},
\eeq
so that we have
\beq
F_\star\approx2z\approx\sqrt{\ii\g}\approx(1+\ii)f_\star,\quad
f_\star\approx\sqrt\frac{\g}{2}.
\label{fsweak}
\eeq
As a consequence of~(\ref{fth}),
the extremal points are given by the scaling formula
\beq
F_k\approx\Phi(x)\,f_\star,
\eeq
with
\beq
x=k\sqrt\frac{\g}{2}=kf_\star,
\eeq
while the complex function $\Phi(x)=U(x)+\ii V(x)$ reads
\beq
U(x)=\frac{\sinh 2x+\sin 2x}{\cosh 2x-\cos 2x},\quad
V(x)=\frac{\sinh 2x-\sin 2x}{\cosh 2x-\cos 2x}.
\eeq
The function $\Phi(x)$ describes the
spiral formed by the extremal points in the continuum limit.
We have $U(\infty)=V(\infty)=1$, and so $\Phi(\infty)=1+\ii$.

The function $U(x)$, describing the real parts of the extremal points,
reaches its smallest minimum for $x=\pi/2$.
We thus obtain the following predictions
in the regime of a weak trapping strength:
the index $k(\g)$ diverges as
\beq
k(\g)\approx\frac{\pi}{\sqrt{2\g}},
\eeq
while the factor $f$ scales as
\beq
f\approx\tau\sqrt\frac{\g}{2}\approx\tau f_\star,
\label{fweak}
\eeq
with
\beq
\tau=U\left(\frac{\pi}{2}\right)=\tanh\frac{\pi}{2}\approx0.917152.
\eeq

\begin{figure}[!ht]
\begin{center}
\includegraphics[angle=-90,width=.55\linewidth]{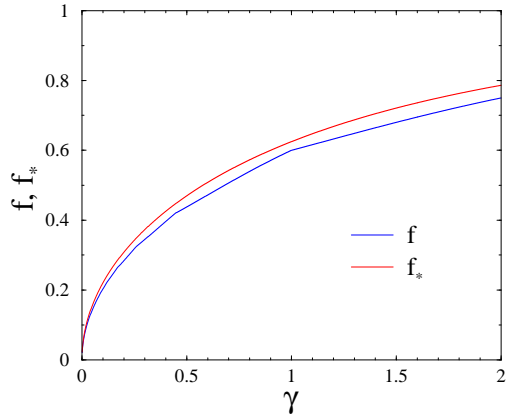}
\caption{\small
Plot of the factor $f$ (see~(\ref{fres})) (lower blue curve)
and of the real part $f_\star$ of the stable fixed point $F_\star$ (see~(\ref{fstar}))
(upper red curve), against $\g$
(Color online).}
\label{fs}
\end{center}
\end{figure}

\begin{figure}[!ht]
\begin{center}
\includegraphics[angle=-90,width=.55\linewidth]{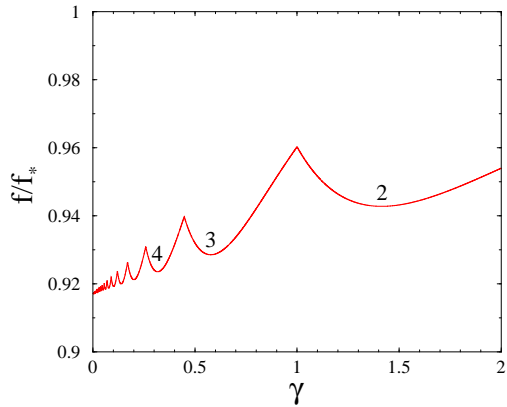}
\caption{\small
Plot of the ratio $f/f_\star$ against $\g$.
Numbers: values of the integer $k(\g)$
labeling the first few regions where this integer is constant
(Color online).}
\label{fratio}
\end{center}
\end{figure}

Figure~\ref{fs} shows a plot of $f$ and $f_\star$ against $\g$.
Both quantities are close to each other,
with $f$ being slightly smaller than $f_\star$.
They exhibit the same overall monotonous dependence on $\g$,
starting as $\sqrt{\g}$ (see~(\ref{fsweak}) and~(\ref{fweak})),
and converging to unity with $1/\g^2$ corrections
(see~(\ref{fstrong}) and~(\ref{fsstrong})).
Figure~\ref{fratio} shows a plot of the ratio $f/f_\star$ against $\g$.
This ratio is slightly smaller than unity,
and it varies between its $\g\to0$ and $\g\to\infty$ limits,
i.e., respectively, $\tau=0.917152$ and unity.
The cusps, corresponding to the points where the integer $k(\g)$ jumps by one unit,
are more clearly visible than in Figure~\ref{fs}.
The first few regions where the integer $k(\g)$ is constant
are labelled by the corresponding values of this integer.

In the limit of an infinite system,
the asymptotic decay of the survival probability $\P(t)$
of the quantum particle
can be estimated by superposing the contributions of Lifshitz regions of all sizes.
For a given (large) size $N$,
the lowest decay rate $\lambda_\min$ is given by~(\ref{lmin}).
Working with exponential accuracy and going to a continuum approximation, we get
\beq
\P(t)\sim\int_0^\infty\exp\left(-\,\frac{8\pi^2ft}{N^3\g}-\abs{\ln(1-c)}\!N\right)\d N.
\eeq
Finally, evaluating the integral by the saddle-point method, we obtain
\beq
\P(t)\sim\exp
\left(-\,\frac{4}{3}\left(\frac{24\pi^2ft}{\g}\right)^{1/4}\abs{\ln(1-c)}^{3/4}\right).
\eeq
We have thus derived the stretched exponential law of the survival probability announced
in~(\ref{pq}) for $d=1$, with
\beq
B_1=\frac{4}{3}\left(\frac{24\pi^2f}{\g}\right)^{1/4}\abs{\ln(1-c)}^{3/4}.
\label{b1res}
\eeq
This exact result involves the non-trivial factor $f$,
defined in~(\ref{fres}) and plotted in Figure~\ref{fs},
which only depends on the trapping strength $\g$.

\section{Higher-dimensional disordered systems}
\label{highd}

This section is devoted to an extension
of the results of Section~\ref{1d} to higher-dimensional disordered systems.

\subsection{A classical particle}
\label{highdc}

We start with a reminder on a classical particle diffusing
on a $d$-dimensional lattice, chosen to be hypercubic for simplicity,
in the presence of a random distribution of traps with density $c$.
We again assume that all traps have the same strength $\g$.
We are thus led to consider the stationary equation
\beq
-E\,p_\n=\sum_{\m(\n)}(p_\m-p_\n)-\g\eta_\n\,p_\n,
\label{chigh}
\eeq
where $\m(\n)$ are the $2d$ neighbors of site $\n$.

In analogy with the one-dimensional situation,
investigated in Section~\ref{1dc},
the smallest eigenvalues of~(\ref{chigh}) correspond
to the lowest modes living in {\it Lifshitz spheres},\footnote{We keep with
the long tradition in mathematical physics~\cite{lgp,fp}
of using the word {\it sphere},
even though the word {\it ball} would be more appropriate
to describe the volume delimited by a sphere.}
i.e., large and almost spherical regions of the lattice which are free of traps.
Consider such a sphere of radius $R\gg1$,
and translate the co-ordinate system so that the sphere is centered at the origin.
The probability of occurrence of such a large trap-free sphere
per unit volume scales as
\beq
P(R)\sim(1-c)^{\omega_d R^d},
\label{spro}
\eeq
where
\beq
\omega_d=\frac{\pi^{d/2}}{\Gamma\left(\frac{d+2}{2}\right)}
\eeq
is the volume of the unit $d$-dimensional sphere.

As we are only interested in the lowest mode
of the discrete difference equation~(\ref{chigh}),
it is legitimate to replace it by its continuum approximation,
\beq
-Ep=\Delta p,
\label{lap}
\eeq
where $\Delta$ is the Laplace operator inside the sphere,
with Dirichlet boundary conditions.
The lowest mode of the above equation is spherically symmetric and reads
\beq
p(r)=L_d(qr),
\label{sp}
\eeq
where $q=\sqrt{E}$ is the wavevector,
while the scaling function $L_d$, normalized to be unity at the center, reads
\beq
L_d(x)=\Gamma\left(\frac{d+2}{2}\right)\left(\frac{2}{x}\right)^{(d-2)/2}J_{(d-2)/2}(x),
\label{sld}
\eeq
where $J_{(d-2)/2}$ is a Bessel function.
The boundary condition $p(R)=0$ yields
\beq
q=\frac{j_d}{R},
\label{sq}
\eeq
where $j_d$ is the first zero of $J_{(d-2)/2}$.
Table~\ref{spheres} gives the value of $\omega_d$, $L_d$ and~$j_d$
in one, two, and three dimensions.

\begin{table}[!ht]
\begin{center}
\begin{tabular}{|c||c|c|c|}
\hline
$d$ & $\omega_d$ & $L_d(x)$ & $j_d$\\
\hline
1 & 2 & $\cos x$ & $\pi/2$\\
2 & $\pi$ & $J_0(x)$ & 2.404825\\
3 & $4\pi/3$ & $(\sin x)/x$ & $\pi$\\
\hline
\end{tabular}
\end{center}
\caption{\small Values of various characteristics of Lifshitz spheres
in one, two, and three dimensions:
$\omega_d$ is the volume of the unit sphere,
$L_d(x)$ is the scaling function of the lowest mode
of the Laplace-Dirichlet equation in the sphere,
$j_d$ is the first zero of the latter function.}
\label{spheres}
\end{table}

The asymptotic decay of the survival probability $P(t)$
can again be estimated by superposing the contributions of Lifshitz spheres of all sizes.
Working with exponential accuracy and going to a continuum approximation, we get
\beq
P(t)\sim\int_0^\infty\exp\left(-\,\frac{j_d^2\,t}{R^2}
-\omega_d\abs{\ln(1-c)}R^d\right)R^{d-1}\d R.
\eeq
Evaluating the integral by the saddle-point method, we obtain
\beq
P(t)\sim\exp\left(-\,C_d\abs{\ln(1-c)}^{2/(d+2)}t^{d/(d+2)}\right).
\eeq
This is the stretched exponential law of the survival probability announced
in~(\ref{pc}),
with
\beq
A_d=C_d\abs{\ln(1-c)}^{2/(d+2)},
\label{adres}
\eeq
where the exact prefactor reads
\beq
C_d=\frac{d+2}{d}\left(\frac{d\omega_d\,j_d^d}{2}\right)^{2/(d+2)}.
\eeq

\subsection{A quantum particle}
\label{highdq}

Let us now consider a quantum particle on a $d$-dimensional lattice,
in the presence of a random distribution of traps with strength $\g$
and density~$c$.
We are thus led to study the stationary tight-binding equation
\beq
\eps\,p_\n=\sum_{\m(\n)}p_\m-\ii\g\eta_\n\,p_\n.
\label{qhigh}
\eeq

On a very large but finite sample,
the survival probability decays exponentially fast, as
\beq
\P(t)\sim\e^{-\lambda t},
\eeq
where the decay rate $\lambda$ is dictated by the eigenvalue of~(\ref{qhigh})
whose imaginary part is the smallest, i.e.,
\beq
\lambda=-2\Im\eps_1.
\label{highl}
\eeq

It is to be expected that $\eps_1$ is close to the band edge
(i.e., $\eps_\max=2d$ on the hypercubic lattice).
This legitimates the use of a continuum description,
and therefore the analysis in terms of Lifshitz spheres,
recalled in Section~\ref{highdc} within the classical framework.
More precisely, setting
\beq
\eps=2d-q^2,
\eeq
the lowest mode in a sphere of large radius ($R\gg1$) is still given by~(\ref{sp}),
and so the real part of the wavevector $q$ is still given by~(\ref{sq}).
We thus obtain
\beq
\Re\eps\approx2d-\frac{j_d^2}{R^2}.
\label{rad}
\eeq

The decay rate $\lambda$ is given by the imaginary part of $\eps$
(see~(\ref{highl})), which is typically much smaller than $1/R^2$,
and whose determination is accordingly more delicate.
In the one-dimensional situation,
the Riccati formalism allowed us to derive the quantitative result~(\ref{lmin}),
including the exact expression~(\ref{fres}) for the factor~$f$.
In the present higher-dimensional situation, no such quantitative approach is available.

An approximation which can be dealt with by analytical means
consists in considering a trap-free spherical cavity of radius $R$
embedded in an infinite space
with a uniform optical potential $\ii W$, with $W=c\g$~\cite{P3}.
For a large enough cavity, and with the present notations,
this model yields the prediction
\beq
\lambda_{\rm cavity}\approx\frac{2j_d^2\Lambda}{R^3}
\approx\frac{4j_d^2}{R^3\sqrt{2c\g}},
\label{cavity}
\eeq
where $\Lambda=\sqrt{2/W}$ is the penetration depth of a particle
with very long wavelength ($q\to0$) into the medium endowed with the optical potential.
The proportionality of the penetration depth
(and hence of the decay rate) to $1/\sqrt{\g}$
is an artifact of the continuum description of the trapping medium.
A $1/\g$ behavior is indeed expected to universally hold
in the presence of discrete traps,
in accord with all the results obtained so far.
We are thus led to make the following Ansatz for the lowest decay rate~$\lambda_\min$
associated with a large Lifshitz sphere of fixed radius $R$:
\beq
\lambda_\min\approx\frac{\phi}{R^3\g}.
\label{anphi}
\eeq
The above Ansatz is meant to capture the leading dependence of the decay rate
on $R$ and $\g$ in the regime where these parameters are large.
The numerator $\phi$ is a higher-dimensional analogue
of the factor $f$ which enters the exact one-dimensional result~(\ref{lmin}).
It is assumed to be of order unity,
and possibly depends on the trap density~$c$.

The asymptotic decay of the survival probability can now be estimated~as
\beq
\P(t)\sim\int_0^\infty\exp\left(-\,\frac{\phi t}{R^3\g}
-\omega_d\abs{\ln(1-c)}R^d\right)R^{d-1}\d R.
\eeq
Evaluating the integral by the saddle-point method, we obtain
\beq
\P(t)\sim\exp\left(-\,Q_d\abs{\ln(1-c)}^{3/(d+3)}\left(\frac{\phi t}{\g}\right)^{d/(d+3)}\right).
\label{shigh}
\eeq
This is the stretched exponential law announced in~(\ref{pq}),
with
\beq
B_d=Q_d\left(\frac{\phi}{\g}\right)^{d/(d+3)}\abs{\ln(1-c)}^{3/(d+3)},
\label{bdres}
\eeq
where the prefactor reads
\beq
Q_d=\frac{d+3}{d}\left(\frac{d\,\omega_d}{3}\right)^{3/(d+3)}.
\eeq
The only part of the result~(\ref{bdres}) which is not known analytically
is the factor~$\phi$,
which has been introduced in the Ansatz~(\ref{anphi}).

\section{Concluding remarks}
\label{concl}

In this work we have investigated the survival probabilities
of a classical and of a quantum particle
in the presence of traps in a whole range of finite or infinite geometries.
Our main goal was to present a thorough comparison
of the temporal behavior of the classical and quantum survival probabilities,
in order to better grasp the qualitative differences between both situations.
This systematic kind of approach had not been adopted so far,
in spite of the appreciable number of published works on the trapping of continuous-time
quantum walks~\cite{P1,P2,P3,ParrisJSP,Lindenberg,Huber,Silbey,KNS,MBA,MPB,AMB,Ag,MulkenRep,GAS,Gonulol,Anish}.

It is worth recalling the leitmotiv of the present study:
static traps are far less efficient
to absorb quantum particles than classical ones.
Free quantum particles indeed benefit from interference effects
in order to efficiently avoid any particular site,
much more easily than diffusing classical particles.
This qualitative difference manifests itself in all the geometries
that we have considered in this work.
On the chain with a single trap, a classical walker is eventually absorbed with certainty,
while a quantum one may escape forever with a rather high probability $\Pi_\infty$.
This asymptotic probability bears a non-monotonic dependence on the trapping strength,
and goes paradoxically to unity in the nominal strong-trapping regime.
On a finite ring of $N$ sites with a single trap,
we have the extra feature that characteristic times scale differently.
In the classical case the relevant time scale is the diffusive one, scaling as $N^2$,
whereas the much longer quantum absorption time grows as $N^3\g$.
Furthermore, for some configurations of several traps on a ring,
the quantum particle may have a non-zero asymptotic survival probability,
owing to the existence of avoiding modes.
In higher dimensions, virtually all questions concerning these
avoiding modes are entirely open:
What is the number of these modes, their nature,
their relevance to quantum percolation?
We plan to return to these matters in future work.

In the presence of a random distribution of traps with fixed density,
both the classical and the quantum survival probabilities
were known to obey a stretched exponential asymptotic decay,
albeit with different exponents,
i.e., respectively $d/(d+2)$ and $d/(d+3)$ (see~(\ref{pc}),~(\ref{pq})).
One of our main goals was to determine the amplitude $B_d$
of the quantum survival probability,
and chiefly its dependence on the trap density $c$ and on the trapping strength $\g$.
This task is far more involved than the evaluation of the classical amplitude~$A_d$.
Indeed, at variance with the classical case,
the decay rate of the quantum particle depends to leading order
on the whole environment.

In the one-dimensional situation,
the strong contrast between the classical and the quantum situations
can be illustrated explicitly.
In the classical case,
in order to evaluate the amplitude $A_1$ (see~(\ref{a1res})),
one has to perform a single averaging over the size $N$ of the Lifshitz region.
In the quantum case, however,
in order to evaluate the amplitude $B_1$ (see~(\ref{b1res})),
one is naturally led to perform a double averaging,
first on the environment at fixed $N$, obtaining thus the non-trivial factor $f$ exactly,
and then over the size $N$.
In higher dimension, however, the first averaging cannot be performed exactly.
We have replaced it by a motivated but heuristic Ansatz.
It seems plausible that evaluating the amplitude $B_d$
by analytical means is just as difficult as evaluating the survival probability
of the classical situation throughout the Lifshitz regime,
beyond its leading term~\cite{inst1,inst2,inst3,inst4,NLEPL,inst5,inst6}.


\begin{thebibliography}{99}

\bibitem{mw} Montroll, E.W., Weiss, G.H.: Random Walks on Lattices. II. J. Math. Phys. {\bf 6}, 167-181 (1965)

\bibitem{BV} Balagurov, B.Ya., Vaks, V.G.: Random walk of a particle in a lattice with traps. J.E.T.P. {\bf 38}, 968-971 (1974)

\bibitem{Varadhan1} Donsker, M., Varadhan, S.R.S.: Asymptotics for the Wiener sausage. Commun. Pure Appl. Math. {\bf 28}, 525-565 (1975)

\bibitem{Varadhan2} Donsker, M., Varadhan, S.R.S.: On the number of distinct sites visited by a random walk. Commun. Pure Appl. Math. {\bf 32}, 721-747 (1979)

\bibitem{Grassberger} Grassberger, P., Procaccia, I.: The long time properties of diffusion in a medium with static traps. J. Chem. Phys. {\bf 77}, 6281-6284 (1982)

\bibitem{Wilczek} Toussaint, D., Wilczek, F.: Particle-antiparticle annihilation in diffusive motion. J. Chem. Phys. {\bf 78}, 2642-2647 (1983)

\bibitem{Klafter} Klafter, J., Zumofen, G., Blumen, A.: Long-time properties of trapping on fractals. J. Phys. (Paris) {\bf 45}, L49-L56 (1984)

\bibitem{Lebowitz1} Bramson, M., Lebowitz, J.L.: Asymptotic behavior of densities in diffusion-dominated annihilation reactions. Phys. Rev. Lett. {\bf 61}, 2397-2400 (1988)

\bibitem{Lebowitz2} Bramson, M., Lebowitz, J.L.: Asymptotic behavior of densities in diffusion-dominated annihilation reactions (erratum). Phys. Rev. Lett. {\bf 62}, 694-694 (1989)

\bibitem{BRW} Ben-Naim, E., Redner, S., Weiss, G.H.: Partial absorption and 'virtual' traps. J. Stat. Phys. {\bf 71}, 75-85 (1993)

\bibitem{hk} Haus, J.W., Kehr, K.W.: Diffusion in regular and disordered lattices. Phys. Rep. {\bf 150}, 263-406 (1987)

\bibitem{DanyRW} Havlin, S., ben-Avraham, D.: Diffusion in disordered media. Adv. Phys. {\bf 36}, 695-798 (1987)

\bibitem{PaulKBook} Krapivsky, P.L., Redner, S., Ben-Naim, E.: A Kinetic View of Statistical Physics. Cambridge University Press, Cambridge (2010)

\bibitem{lif} Lifshitz, I.M.: The energy spectrum of disordered systems. Adv. Phys. {\bf 13}, 483-536 (1964)

\bibitem{lgp} Lifshitz, I.M., Gredeskul, S.A., Pastur, L.A.: Introduction to the Theory of Disordered Systems. Wiley, New York (1988)

\bibitem{fp} Pastur, L., Figotin, A.: Spectra of Random and Almost-Periodic Operators. Springer, Berlin (1992)

\bibitem{Rosenstock} Rosenstock, H.B.: Random walks on lattices with traps. J. Math. Phys. {\bf 11}, 487-490 (1970)

\bibitem{Barkema} Barkema, G.T., Biswas, P., van Beijeren, H.: Diffusion with random distribution of static traps. Phys. Rev. Lett. {\bf 87}, 170601 (2001)

\bibitem{inst1} Friedberg, R., Luttinger, J.M.: Density of electronic energy levels in disordered systems. Phys. Rev. B {\bf 12}, 4460-4474 (1975)

\bibitem{inst2} Cardy, J.L.: Electron localisation in disordered systems and classical solutions in Ginzburg-Landau field theory. J. Phys. C {\bf 11}, L321-L327 (1978)

\bibitem{inst3} Luttinger, J.M., Tao, R.: Electronic density of levels in a disordered system. Ann. Phys. {\bf 145}, 185-203 (1983)

\bibitem{inst4} Lubensky, T.C.: Fluctuations in random walks with random traps. Phys. Rev. A {\bf 30}, 2657-2665 (1984)

\bibitem{NLEPL} Nieuwenhuizen, T.M., Luck, J.M.: Singularities in spectra of disordered systems: an instanton approach for arbitrary dimension and randomness. Europhys. Lett. {\bf 9}, 407-413 (1989)

\bibitem{inst5} Nieuwenhuizen, T.M.: Trapping and Lifshitz tails in random media, self-attracting polymers, and the number of distinct sites visited: a renormalized instanton approach in three dimensions. Phys. Rev. Lett. {\bf 62}, 357-360 (1989)

\bibitem{inst6} Zee, A., Affleck, I.: Hopping between random locations: spectrum and instanton. J. Phys. Cond. Matt. {\bf 12}, 8863-8873 (2000)

\bibitem{P1} Parris, P.E.: One-dimensional trapping kinetics at zero temperature. Phys. Rev. Lett. {\bf 62}, 1392-1395 (1989)

\bibitem{P2} Parris, P.E.: One-dimensional quantum transport in the presence of traps. Phys. Rev. B {\bf 40}, 4928-4937 (1989)

\bibitem{P3} Edwards, J.W., Parris, P.E.: Quantum transport in the presence of random traps. Phys. Rev. B {\bf 40}, 8045-8048 (1989)

\bibitem{ParrisJSP} Parris, P.E.: Quantum and stochastic aspects of low-temperature trapping and reaction dynamics. J. Stat. Phys. {\bf 65}, 1161-1172 (1991)

\bibitem{RMP} Pearlstein, R.M.: Impurity quenching of molecular excitons. I. Kinetic comparison of F\"orster—Dexter and slowly quenched Frenkel excitons in linear chains. J. Chem. Phys. {\bf 56}, 2431-2442 (1972)

\bibitem{Lindenberg} Hemenger, R.P., Lakatos-Lindenberg, K., Pearlstein, R.M.: Impurity quenching of molecular excitons. III. Partially coherent excitons in linear chains. J. Chem. Phys. {\bf 60}, 3271-3277 (1974)

\bibitem{Huber} Huber, D.L.: Decay of the k⃗=0 exciton mode at a finite trap concentration. Phys. Rev. B {\bf 24}, 1083-1086 (1981)

\bibitem{Silbey} Benk, H., Silbey, R.: Exciton-phonon scattering and exciton trapping in one-dimensional exciton systems. J. Chem. Phys. {\bf 79}, 3487-3495 (1983)

\bibitem{ADZ} Aharonov, Y., Davidovich, L., Zagury, N.: Quantum random walks. Phys. Rev. A {\bf 48}, 1687-1690 (1993)

\bibitem{FG} Farhi, E., Gutmann, S.: Quantum computation and decision trees. Phys. Rev. A {\bf 58}, 915-928 (1998)

\bibitem{Goldstone} Childs, A.M., Goldstone, J.: Spatial search by quantum walk. Phys. Rev. A {\bf 70}, 022314 (2004)

\bibitem{ChildsPRL} Childs, A.M.: Universal computation by quantum walk. Phys. Rev. Lett. {\bf 102}, 180501 (2009)

\bibitem{AC} Childs, A.M.: On the relationship between continuous- and discrete-time quantum walk. Commun. Math. Phys. {\bf 294}, 581-603 (2010)

\bibitem{Kempe} Kempe, J.: Quantum random walks: an introductory overview. Contemp. Phys. {\bf 44}, 307-327 (2003)

\bibitem{Ambainis} Ambainis, A.: Quantum walks and their algorithmic applications. Int. J. Quant. Inf. {\bf 1}, 507-518 (2003)

\bibitem{perets} Perets, H.B., Lahini, Y., Pozzi, F., Sorel, M., Morandotti, R., Silberberg, Y.: Realization of quantum walks with negligible decoherence in waveguide lattices. Phys. Rev. Lett. {\bf 100}, 170506 (2008)

\bibitem{peruzzo} Peruzzo, A., Lobino, M., Matthews, J.C.F., Matsuda, N., Politi, A., Poulios, K., Zhou, X.Q., Lahini, Y., Ismail, N., W\"orhoff, K., Bromberg, Y., Silberberg, Y., Thompson, M.G., OBrien, J.L.: Quantum walks of correlated photons. Science {\bf 329}, 1500-1503 (2010)

\bibitem{NK} Konno, N.: Quantum random walks in one dimension. Quant. Inf. Processing {\bf 1}, 345-354 (2002)

\bibitem{MBT} Mackay, T.D., Bartlett, S.D., Stephenson, L.T., Sanders, B.C.: Quantum walks in higher dimensions. J. Phys. A {\bf 35}, 2745-2753 (2002)

\bibitem{TFM} Tregenna, B., Flanagan, W., Maile, R., Kendon, V.: Controlling discrete quantum walks: coins and initial states. New J. Phys. {\bf 5}, 83 (2003)

\bibitem{TL} de Toro Arias, S., Luck, J.M.: Anomalous dynamical scaling and bifractality in the one-dimensional Anderson model. J. Phys. A {\bf 31}, 7699-7717 (1998)

\bibitem{benAvraham} ben-Avraham, D., Bollt, E.M., Tamon, C.: One-dimensional continuous-time quantum walks. Quant. Inf. Processing {\bf 3}, 295-308 (2004)

\bibitem{Eric} Akkermans, E., Comtet, A., Desbois, J., Montambaux, G., Texier, C.: Spectral determinant on quantum graphs. Ann. Phys. {\bf 284}, 10-51 (2000)

\bibitem{FS} Strauch, F.W.: Connecting the discrete- and continuous-time quantum walks. Phys. Rev. A {\bf 74}, 030301 (2006)

\bibitem{qwrev} Venegas-Andraca, S.E.: Quantum walks: a comprehensive review. Quant. Inf. Processing {\bf 11}, 1015-1106 (2012)

\bibitem{Haroche} Haroche, S., Raimond, J.M.: Exploring the Quantum: Atoms, Cavities, and Photons. Oxford University Press, New York (2006)

\bibitem{KNS} Konno, N., Namiki, T., Soshi, T., Sudbury, A.: Absorption problems for quantum walks in one dimension. J. Phys. A {\bf 36}, 241-253 (2003)

\bibitem{MBA} M\"ulken, O., Blumen, A., Amthor, T., Giese, C., Reetz-Lamour, M., Weidenm\"uller, M.: Survival probabilities in coherent exciton transfer with trapping. Phys. Rev. Lett. {\bf 99}, 090601 (2007)

\bibitem{MPB} M\"ulken, O., Pernice, V., Blumen, A.: Slow excitation trapping in quantum transport with long-range interactions. Phys. Rev. E {\bf 78}, 021115 (2008)

\bibitem{AMB} Agliari, E., M\"ulken, O., Blumen, A.: Continuous-time quantum walks and trapping. Int. J. Bif. Chaos {\bf 20}, 271-279 (2010)

\bibitem{Ag} Agliari, E.: Trapping of continuous-time quantum walks on Erd\"os-R\'enyi graphs. Physica A {\bf 390}, 1853-1860 (2011)

\bibitem{MulkenRep} M\"ulken, O., Blumen, A.: Continuous-time quantum walks: models for coherent transport on complex networks. Phys. Rep. {\bf 502}, 37-87 (2011)

\bibitem{GAS} G\"on\"ulol, M., Aydiner, E., Shikano, Y., M\"ustecaplioglu, \"O.E.: Survival probability in a one-dimensional quantum walk on a trapped lattice. New J. Phys. {\bf 13}, 033037 (2011)

\bibitem{Gonulol} Ampadu, C., G\"on\"ulol, M., Aydiner, E.: Survival probability of the quantum walk with phase parameters on the two-dimensional trapped lattice. J. Quant. Inf. Sc. {\bf 3}, 51-56 (2013)

\bibitem{Anish} Anishchenko, A., Blumen, A., M\"ulken, O.: Geometrical aspects of quantum walks on random two-dimensional structures. Phys. Rev. E {\bf 88}, 062126 (2013)

\bibitem{NL} Nieuwenhuizen, T.M., Luck, J.M.: Lifshitz singularities in random harmonic chains: periodic amplitudes near the band edge and near special frequencies. J. Stat. Phys. {\bf 48}, 393-424 (1987)

\bibitem{JMLivre} Luck, J.M.: Syst\`emes d\'esordonn\'es unidimensionnels. Collection Al\'ea Saclay (1992)

\bibitem{Selsto1} Selst{\o}, S.: Formulae for partial widths derived from the Lindblad equation. Phys. Rev. A {\bf 85}, 062518 (2012)

\bibitem{Selsto2} Selst{\o}, S., Kvaal, S.: Absorbing boundary conditions for dynamical many-body quantum systems. J. Phys. B {\bf 43}, 065004 (2010)

\bibitem{Diaz} Diaz-Torres, A.: Absence of decoherence in the complex-potential approach to nuclear scattering. Phys. Rev. C {\bf 81}, 041603(R) (2010)

\bibitem{gr} Gradshteyn, I.S., Ryzhik, I.M.: Table of Integrals, Series, and Products. Academic Press, New York (1965)

\bibitem{HOZ} Holz, D.E., Orland, H., Zee, A.: On the remarkable spectrum of a non-Hermitian random matrix model. J. Phys. A. {\bf 36}, 3385-3400 (2003)

\end{thebibliography}
\end{document}